\documentclass[aps,prb,twocolumn,amsmath,amssymb]{revtex4-2}
\usepackage{graphics}
\usepackage{epsfig}
\usepackage{epsf,epic}
\usepackage{dcolumn}
\usepackage{color}
\usepackage{hyperref}
\usepackage{marvosym} 
\usepackage{amsmath}
\usepackage{amssymb}
\usepackage{amsfonts}
\usepackage{wrapfig}
\usepackage{pstricks}
\usepackage{multirow}
\graphicspath{{\figures}}
\usepackage{bm}
\usepackage{dcolumn}
\usepackage{tikz}
\usepackage{amssymb}
\usepackage{psfrag}
\usepackage{textcomp}
\usepackage{pdfsync}
\usepackage{amsmath}
\usepackage{amsfonts}
\usepackage{graphicx,bm,epstopdf}
\usepackage[FIGTOPCAP]{subfigure}
\usepackage{upgreek}
\usepackage[euler]{textgreek}
\usepackage{natbib}
\usepackage{threeparttable}
\usepackage[normalem]{ulem}
\usepackage{amsmath}
\usepackage{amssymb}
\usepackage{graphicx}
\usepackage[]{hyperref}
\hypersetup{
    colorlinks=true,
    citecolor=blue,
    linkcolor=blue,
    urlcolor=blue,
    }
\usepackage{float}
\usepackage{placeins}

\begin{document}
\title{Roles of Electron-Magnon Cross Diffusion in Unidirectional Magnetoresistance of Metallic Magnetic Bilayers}

\author{Shashank Gupta}
\author{Steven S.-L. Zhang}
\thanks{Contact author: shulei.zhang@case.edu}

\affiliation{Department of Physics, Case Western Reserve University, 10900 Euclid Avenue, Cleveland, Ohio 44106-7079, USA}

\begin{abstract}
  Unidirectional magnetoresistance (UMR) in metallic bilayers arises from nonlinear spin-charge transport mediated by broken time-reversal and inversion symmetries, yet the role of magnons remains unsettled. We develop a theoretical framework that incorporates coupled electron-magnon dynamics, revealing cross diffusion and spin-angular-momentum transfer between the two subsystems, which renormalize the characteristic electron and magnon spin-diffusion lengths. We show that nonequilibrium magnons, indirectly excited by the electric field, can suppress UMR by absorbing spin angular momentum from conduction electrons. We also analyze the magnetic-field, thickness, and temperature dependencies and identify distinct features that constitute experimental fingerprints of magnonic contributions to UMR in metallic bilayers, providing qualitative to semiquantitative guidance for elucidating the underlying physical mechanisms.
 \end{abstract}
\maketitle
\section{Introduction}
Unidirectional magnetoresistance (UMR) in metallic magnetic bilayers~\cite{avci2015natphys,Ferguson16APL_UMR-STO(exp)} manifests as a resistance change upon reversal of either the current polarity or the magnetization direction, in fundamental contrast to linear-response effects such as anisotropic magnetoresistance~\cite{McGuire75TMag_AMR,Thompson75TMag_AMR-thin-film,Kobs11PRL_iAMR-PtCoPt,Chien12PRL_Proximity-Pt,clChien13PRB_AMR-FM-NM,Z&Z14JAP_AMR,slzhang15PRB_AMR} and spin Hall magnetoresistance~\cite{Saitoh13PRL_SH-MR,yChen13PRB_theory-SMR,Althammer13PRB_quant-SMR-FI-NM,Hayashi16PRL_SMR-metallic-BL}, which remain invariant under such reversals. From a symmetry perspective, UMR is a nonlinear transport effect that requires simultaneous breaking of time-reversal and inversion symmetries, and is not constrained by Onsager reciprocity relations that apply in the linear-response regime. Metallic bilayers consisting of a ferromagnetic metal (FM) and a nonmagnetic metal (NM) naturally satisfy these conditions: time-reversal symmetry is spontaneously broken for the FM layer, while inversion symmetry is broken by the structural asymmetry of the FM$\vert$NM interface. 

Aside from its fundamental significance, UMR in metallic bilayers provides a compact two-terminal electrical readout mechanism of the magnetization state in spintronic devices, enabling the significant simplification of established spin–orbit torque magnetic random-access memory (SOT-MRAM) architectures~\cite{pnHai19JAP_UMR-MRAM} and opening opportunities for new paradigms such as two-terminal multi-state magnetic memories built from spin-valve cells~\cite{Avci17APL_memory-UMR}.

Despite extensive experimental studies of UMR over the past decade, a complete understanding of its underlying physical mechanisms remains elusive. In metallic bilayers, UMR has generally been attributed to the combined action of current-induced spin accumulation and spin-dependent scattering processes, occurring both at the FM$\vert$NM interface and within the bulk of the ferromagnet~\cite{avci2015natphys,Zhang&Vignale16PRB_USMR,Zhang&Vignale17SPIE_UMR,avci2018prl}, where the spin accumulation originates either from the spin Hall effect in the NM~\cite{DYAKONOV71PLA_spinHall,Hirsch99PRL_SHE,sZhang00PRL_SHE,Vignale10_SHE,aHoffman13IEEE_SHE,Sinova15RMP_SHE} or from the spin-anomalous-Hall effect in the FM layer~\cite{Amin19PRB_spin-current-FM,M&Z22PRB_AH-UMR}. Experimental studies have also evidenced that nonequilibrium magnons, excited by current-induced spin accumulation, can strongly influence UMR, particularly at elevated temperatures~\cite{Ferguson16APL_UMR-STO(exp),tOno17APExpr_UMR-magnon(exp),Demokritov18APL_UMR-magnon(exp), avci2018prl,Gambardelaa25PRL_UMR-magnon(exp),Sanghoon23PRAppl_magnon-UMR,Kim_2019}. However, it remains unclear whether the relevant magnons are of exchange or dipolar origin, and a systematic theoretical framework that formulates the role of nonequilibrium magnons in UMR hosted by metallic bilayers has not yet been established.

Microscopically, current-induced nonequilibrium magnons may modify either the electron relaxation times or the nonequilibrium electron density (more specifically, the electron spin accumulation). In this work, we restrict ourselves to the latter case and develop a theory for nonreciprocal and nonlinear charge transport in metallic magnetic bilayers arising from the interplay between electron and magnon transport. Our approach provides a systematic framework to capture these effects: we treat spin transport mediated by conduction electrons and exchange magnons on an equal footing by solving coupled kinetic equations, supplemented by interfacial boundary conditions that capture the transmutation among electron spin current, magnon accumulation, magnon spin current, and electron spin accumulation.

We find that electron–magnon interactions give rise to cross-diffusion between current-induced nonequilibrium electrons and magnons perpendicular to the layer plane. Because this cross-diffusion conserves the total angular momentum of the composite system, an increase in the nonequilibrium magnon population within the FM layer reduces the electron spin accumulation and consequently suppresses the UMR, and vice versa. When such cross-diffusion plays a dominant role in the UMR, suppressing magnons—either by applying an external magnetic field parallel to the equilibrium magnetization or by lowering the temperature—enhances the UMR. Moreover, the position of the UMR peak as a function of the FM layer thickness becomes temperature dependent, shifting to smaller thicknesses at higher temperatures.

The remainder of the paper is organized as follows. Section~\ref{sec:Formulation} outlines the theoretical framework, deriving coupled electron–magnon diffusion equations and establishing interfacial boundary conditions for spin accumulations and current densities of both electrons and magnons. Section~\ref{sec:R&D} presents calculations of the UMR coefficients for FM$|$NM bilayers, uncovering the roles of magnons and electron–magnon interactions in its generation, and quantifying their contribution as key material and structural parameters are varied. These results reveal clear signatures of magnon involvement in UMR that can be tested experimentally. Section~\ref{sec:Conclusion} summarizes the main findings and outlines future directions for exploring magnon-mediated nonreciprocal and nonlinear charge transport in magnetic heterostructures.

\section{Formulations}
\label{sec:Formulation}

\subsection{Coupled electron-magnon kinetic equations}
We employ a semiclassical description of electron and magnon transport, where quasiparticles are treated as wave packets with distribution functions that evolve according to their respective kinetic equations. Equation~(\ref{eq:KE-electron}) governs the evolution of the distribution of conduction electrons with momentum 
$\mathbf{k}$, spin $\sigma$, and spatial coordinate $\mathbf{r}$ in the FM layer, 
under an applied electric field $\mathbf{E}$. On the left-hand side, the first term, involving the group velocity 
$\mathbf{v}_{\mathbf{k}\sigma} = \tfrac{1}{\hbar}\nabla_{\mathbf{k}}\epsilon_{\mathbf{k}\sigma}$, 
and the second term, associated with the electric field $\mathbf{E}$, 
are the convective and drift terms, respectively. On the right-hand side, $\tau_{\sigma}$ denotes the momentum relaxation time for spin~$\sigma$, 
$\tau_{\uparrow\downarrow}$ the spin-flip relaxation time (arising from spin relaxation processes 
other than electron--magnon coupling), and 
$\overline{f}_{\sigma}$ the $\mathbf{k}$-averaged distribution. 

\begin{widetext}
\begin{subequations}
\begin{align}
\label{eq:KE-electron}
\left[\mathbf{v}_{\mathbf k \sigma} \cdot\nabla_{\mathbf r} - (e/\hbar)\mathbf{E}\cdot\nabla_{\mathbf{k}}\right] f_{\mathbf k\sigma}(\mathbf{r}) &= 
-\frac{f_{\mathbf k\sigma}(\mathbf r) - \overline{f_\sigma}(\mathbf r)}{\tau_\sigma} 
-\frac{f_{\mathbf k\sigma}(\mathbf r) - \overline{f_{-\sigma}}(\mathbf r)}{\tau_{\uparrow\downarrow}} 
+\left[\frac{\partial f_{\mathbf k\sigma}(\mathbf r)}{\partial t}\right]_{\mathrm{em}},\\
\label{eq:KE-magnon}
\mathbf{v}_{\mathbf q } \cdot\nabla_{\mathbf r}  n_{\mathbf q}(\mathbf r) &= -\frac{ n_{\mathbf q}(\mathbf r) -{\overline{n}(\mathbf r)}}{\tau_\mathrm{m}} 
- \frac{n_{\mathbf q}(\mathbf r) - n^0_{\mathbf q}}{ \tau_{\mathrm{th}}} 
+ \left[ \frac{\partial n_{\mathbf q}(\mathbf r)}{\partial t} \right]_{\mathrm{em}}.
\end{align}
\end{subequations}
\end{widetext}

For the collision integral on the right-hand side, scattering processes involving electron--magnon interactions are retained explicitly, while all other mechanisms are described within the relaxation-time approximation, separated into (a) spin-conserving processes, characterized by the momentum relaxation time $\tau_{\sigma}$, and (b) spin-flip processes, characterized by the spin-flip relaxation time $\tau_{\uparrow\downarrow}$. In layered structures with confinement, where spin accumulation develops in the steady state, electrons of each spin species do not relax directly to the global equilibrium distribution,  $f_{\mathbf{k}}^{0}$, which is spatially uniform. Instead, scattering drives the distribution toward the local isotropic component $\overline{f}_\sigma(\mathbf{r})$~\cite{sZhang00PRL_SHE}: for spin-conserving processes this reflects momentum randomization within each spin channel, while for spin-flip processes it describes relaxation between the locally averaged spin populations.

Unlike electrons, magnons are charge neutral and therefore do not acquire a drift term driven directly by an electric field. Accordingly, in the magnon kinetic equation (\ref{eq:KE-magnon}) only the convective term appears on the left-hand side, where the magnon group velocity is $\mathbf{v}_{\mathbf{q}}=\frac{1}{\hbar}\nabla_{\mathbf{q}}\omega_{\mathbf{q}}$ with $\omega_{\mathbf{q}}$ the magnon energy. On the right-hand side of the magnon kinetic equation, the first term, with relaxation time $\tau_{\mathrm{m}}$, describes momentum-relaxing processes that
randomize the propagation direction and relax the distribution toward its $\mathbf{q}$-averaged
value $\overline{n}(\mathbf{r})$~\cite{sZhang12PRL}. The second term, with rate 
$\tau_{\mathrm{th}}^{-1}$, accounts for thermalization toward the global Bose--Einstein 
equilibrium $n_{\mathbf{q}}^0$~\cite{sZhang12PRL}; this contrasts with electrons, where relaxation is toward the local isotropic 
distribution of the opposite spin channel in order to conserve the total number of conduction 
electrons. The distinction reflects the fact that magnons are bosonic excitations whose number 
is not conserved. The last term 
captures electron--magnon interactions, acting as a source or sink of magnons coupled to the electronic subsystem.

\begin{figure}[t]
\includegraphics[width=0.4\textwidth]{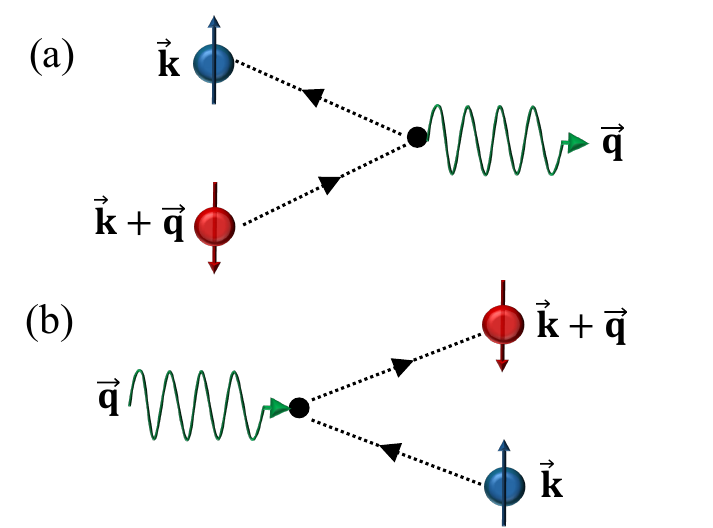}
\caption{Feynman diagrams of electron-magnon scattering processes. (a) Spin-flip scattering of an electron from a spin-down state $(\mathbf{k+q},\downarrow)$ to a spin-up state $(\mathbf{k},\uparrow)$, accompanied by the emission of a magnon with momentum $\mathbf{q}$ that carries an angular momentum quantum of $-\hbar$. (b) Spin-flip scattering of an electron from a spin-up state $(\mathbf{k},\uparrow)$ to a spin-down state $(\mathbf{k+q},\downarrow)$, accompanied by the absorption of a magnon with momentum $\mathbf{q}$ with angular momentum of $-\hbar$.}
 \label{fig:emi-Feynman} 
\end{figure}

So far we have only indicated in general terms that the electron and magnon kinetic equations are coupled through electron–magnon interactions. To make this coupling explicit, we now introduce the second-quantized Hamiltonian for the electron–magnon interaction~\cite{Kasuya56PTP_el-mag-FM,Mannari57PTP_el-mag-FM,Goodings63PR_resistivity-el-magnon,Davis&Liu67PRL_el-mag-intxn}: \begin{equation}\label{eq:em-intxn}
\hat{V}_{\text{em}} = -J_{\mathrm{sd}} \sqrt{\frac{S}{2N}} \sum_{\mathbf {k,q}} 
\big(a_{\mathbf q}^\dagger c_{\mathbf k\uparrow}^\dagger c_{\mathbf k+ \mathbf q\downarrow} +
a_\mathbf q c_{\mathbf k+ \mathbf q\downarrow}^\dagger c_{\mathbf k\uparrow}\big)\,.
\end{equation}
Here $J_{\mathrm{sd}}$ denotes the exchange coupling constant between itinerant electron spins and localized magnetic moments, $N$ the number of atomic sites, $S$ the spin per site, 
$a_{\mathbf{q}}^\dagger \,(a_{\mathbf{q}})$ the creation (annihilation) operators for magnons, 
and $c_{\mathbf{k}\sigma}^\dagger \,(c_{\mathbf{k}\sigma})$ the creation (annihilation) operators 
for electrons with spin $\sigma = \uparrow,\downarrow$. This coupling mediates electron spin flips accompanied by magnon creation or annihilation, thereby transferring spin angular momentum between the electronic and magnonic subsystems while conserving the total. At the same time, momentum is exchanged—when bulk disorder is sufficiently weak to preserve momentum correlations—thereby modifying the distribution of nonequilibrium magnons. The conservation of spin angular momentum and linear momentum in the two coexisting spin-flip electron–magnon scattering processes is illustrated by the Feynman diagrams in Fig.~\ref{fig:emi-Feynman}. These conservation laws underpin the electrically driven generation of magnon currents and accumulations discussed in a moment. 

The full expressions for the electron--magnon collision integrals are obtained by evaluating the 
second-quantized interaction within the 
Born approximation using Fermi's golden rule. This standard procedure yields the explicit collision integrals given below~\cite{Mannari57PTP_el-mag-FM,Goodings63PR_resistivity-el-magnon}:

\begin{subequations}
\label{eq:collision-terms}
\begin{align}
\label{eq:coll-el-up}
\left[\frac{\partial f_{\mathbf k\uparrow}}{\partial t}\right]_{\mathrm{em}}
&= \frac{\pi J_{\mathrm{sd}}^{2} S}{\hbar N}
\sum_{\mathbf q} \delta(\epsilon_{\mathbf k} + \omega_{\mathbf q} - \epsilon_{\mathbf{k+q}}) \notag\\
& \times
\Bigl[(1 - f_{\mathbf k\uparrow}) f_{\mathbf{k+q}\downarrow} (1 + n_{\mathbf q})
- (1 - f_{\mathbf{k+q}\downarrow}) f_{\mathbf k\uparrow} n_{\mathbf q}\Bigr], \\[6pt]
\label{eq:coll-el-dn}
\left[\frac{\partial f_{\mathbf k\downarrow}}{\partial t}\right]_{\mathrm{em}}
&= \frac{\pi J_{\mathrm{sd}}^{2} S}{\hbar N}
\sum_{\mathbf q} \delta(\epsilon_{\mathbf k} - \omega_{\mathbf q} - \epsilon_{\mathbf{k-q}}) \notag\\
& \times
\Bigl[(1 - f_{\mathbf{k}\downarrow}) f_{\mathbf{k-q}\uparrow} n_{\mathbf q}
- (1 - f_{\mathbf{k-q}\uparrow}) f_{\mathbf k\downarrow} (1 + n_{\mathbf q})\Bigr], \\[6pt]
\label{eq:coll-magnon}
\left[\frac{\partial n_{\mathbf q}}{\partial t}\right]_{\mathrm{em}}
&= \frac{\pi J_{\mathrm{sd}}^{2} S}{\hbar N}
\sum_{\mathbf k} \delta(\epsilon_{\mathbf k} + \omega_{\mathbf q} - \epsilon_{\mathbf{k+q}}) \notag\\
& \times
\Bigl[(1 - f_{\mathbf{k}\uparrow}) f_{\mathbf{k+q}\downarrow} (1 + n_{\mathbf q})
- (1 - f_{\mathbf{k+q}\downarrow}) f_{\mathbf k\uparrow} n_{\mathbf q}\Bigr].
\end{align}
\end{subequations}

The electron and magnon distribution functions herein follow from ensemble averages in the Heisenberg 
picture, $f_{\mathbf{k}\sigma}=\langle c_{\mathbf{k}\sigma}^\dagger c_{\mathbf{k}\sigma}\rangle$ 
(with $\sigma=\uparrow,\downarrow$) and 
$n_{\mathbf{q}}=\langle a_{\mathbf{q}}^\dagger a_{\mathbf{q}}\rangle$. These collision integrals incorporate the essential physical ingredients: Pauli blocking factors
$(1-f)$ for electrons and Bose enhancement factors $(n,\,1+n)$ for magnons, ensuring proper
fermionic and bosonic statistics; and delta functions that enforce energy conservation in each
scattering event. These terms
guarantee detailed balance and conservation of total spin angular momentum and linear momentum between the electronic
and magnonic subsystems.

\subsection{Electron-magnon cross-diffusion}

By taking the zeroth and first velocity-weighted moments of the kinetic equations~(\ref{eq:KE-electron}) and (\ref{eq:KE-magnon}), one obtains a set of coupled drift–diffusion equations~\cite{yhCheng17PRB_magnon-el}:
\begin{subequations}
\label{eq:current-continuity and drift-diffusion}
\begin{equation}
\label{eq:current-continuity}
\begin{pmatrix} 
\nabla_{\mathbf r}\cdot\mathbf{j}_{\mathrm{s}}(\mathbf{r}) \\ 
\nabla_{\mathbf r}\cdot\mathbf{j}_{\mathrm{m}}(\mathbf{r}) 
\end{pmatrix}
=
-
\begin{pmatrix} 
\tau_{11}^{-1} & \tau_{12}^{-1} \\ 
\tau_{21}^{-1} & \tau_{22}^{-1} 
\end{pmatrix}
\begin{pmatrix} 
\delta n_{\mathrm{s}}(\mathbf{r}) \\ 
\delta n_{\mathrm{m}}(\mathbf{r}) 
\end{pmatrix},
\end{equation}
\begin{equation}
\label{eq:drift-diffusion}
\begin{pmatrix} 
\mathbf{j}_{\mathrm{s}}(\mathbf r) \\ 
\mathbf{j}_{\mathrm{m}}(\mathbf r) 
\end{pmatrix}
=\mathbf{E}
\begin{pmatrix} 
P_{\sigma}\sigma_{\mathrm{e}} \\ 
\sigma_{\mathrm{m}} 
\end{pmatrix}-
\begin{pmatrix} 
D_{\mathrm{s}} & -D_{\mathrm{sm}} \\ 
-D_{\mathrm{ms}} & D_{\mathrm{m}} 
\end{pmatrix}
\begin{pmatrix} 
\nabla_{\mathbf r}\delta n_{\mathrm{s}}(\mathbf r) \\ 
\nabla_{\mathbf r}\delta n_{\mathrm{m}}(\mathbf r) 
\end{pmatrix}.
\end{equation}
\end{subequations}

The relevant transport variables are the nonequilibrium spin and magnon densities together with their associated currents. 
The nonequilibrium electron spin density (or spin accumulation) is defined as 
$\delta n_{\mathrm{s}}(\mathbf{r}) = \sum_{\mathbf{k}}[f_{\mathbf{k}\uparrow}(\mathbf{r})-f_{\mathbf{k}\downarrow}(\mathbf{r})]$, 
while the nonequilibrium magnon density (magnon accumulation) is 
$\delta n_{\mathrm{m}}(\mathbf{r}) = \sum_{\mathbf{q}}[n_{\mathbf{q}}(\mathbf{r})-n_{\mathbf{q}}^{0}]$. 
The corresponding current densities, which describe the transport of spin angular momentum carried by electrons and magnons, are given by 
$\mathbf{j}_{\mathrm{s}}(\mathbf{r}) = \sum_{\mathbf{k}}[f_{\mathbf{k}\uparrow}(\mathbf{r})-f_{\mathbf{k}\downarrow}(\mathbf{r})]\mathbf{v}_{\mathbf{k}\sigma}$ 
and 
$\mathbf{j}_{\mathrm{m}}(\mathbf{r}) = \sum_{\mathbf{q}}[n_{\mathbf{q}}(\mathbf{r})-n_{\mathbf{q}}^{0}]\mathbf{v}_{\mathbf{q}}$,
respectively.

Equation~(\ref{eq:current-continuity}) represents coupled continuity relations for the electron and magnon spin densities, with relaxation processes appearing on the right-hand side.
The diagonal terms describe intra-subsystem relaxation: $\tau_{11}^{-1}$ characterizes the decay of electron spin accumulation within the electronic channel due to spin-flip mechanisms that do not involve \emph{nonequilibrium} magnons (e.g., Elliott--Yafet–type spin--orbit processes~\cite{Elliott54PR_SOC-spin-relax,Zutic04RMP} and, if included phenomenologically, scattering off a thermal magnon bath treated at equilibrium~\cite{Fert69JPhysC_two-current-magnon}), while $\tau_{22}^{-1}$ accounts for magnon relaxation within the magnon channel.
The off-diagonal terms encode electron--magnon interconversion: $\tau_{21}^{-1}$ corresponds to the transfer of electron spin accumulation into magnon accumulation through magnon emission, whereas $\tau_{12}^{-1}$ describes the reverse process, in which magnons are absorbed to generate an electronic spin imbalance.
These off-diagonal couplings ensure conservation of the total spin angular momentum of the composite electron–magnon system. It can be shown that in the absence of electron-magnon scattering (i.e., $J_{\mathrm{sd}}\to 0$), the continuity equations for electrons and magnons reduce to their uncoupled form~\cite{Fert&Valet93PRB_cpp-GMR,sZhang12PRL,ZZ12PRB_spin-convertance}:
\begin{subequations}\label{eq:uc-continuity-eqs}
\begin{align}
\nabla_{\mathbf r}\cdot\mathbf{j}_{\mathrm{s}}(\mathbf{r})+ \frac{2\delta n_{\mathrm{s}}(\mathbf{r})}{\tau_{\uparrow\downarrow}}&=0\,, \\ 
\nabla_{\mathbf r}\cdot\mathbf{j}_{\mathrm{m}}(\mathbf{r})+\frac{\delta n_{\mathrm{m}}(\mathbf{r})}{\tau_\mathrm{th}}&=0\,.
\end{align}
\end{subequations}

The terms proportional to $\mathbf{E}$ in Eq.~(\ref{eq:drift-diffusion}) represent the generalized Ohm's laws, describing drift spin currents of electrons and magnons  driven by the external electric field. 
For electrons, the term $P_{\sigma}\sigma_{\mathrm{e}}\mathbf{E}$ describes the spin-polarized component of the charge current, with $\sigma_0$ the electron Drude conductivity and $P_{\sigma}(\equiv \frac{\sigma_{\uparrow}-\sigma_{\downarrow}}{\sigma_{\uparrow}+\sigma_{\downarrow}})$ the conductivity spin polarization. 
For magnons, the term $\sigma_\mathrm{m}\mathbf{E}$ denotes a magnon spin current driven electrically through electron--magnon scattering processes~\cite{yhCheng17PRB_magnon-el}, which convert part of the electron spin current into a nonequilibrium flow of magnons by transferring momentum to the magnon subsystem, provided that momentum correlations are preserved in the presence of weak disorder. 

The terms in Eq.~(\ref{eq:drift-diffusion}) associated with the density gradients represent the generalized Fick’s law, capturing diffusive spin currents of electrons and magnons driven by nonequilibrium density gradients.
The coefficients $D_{\mathrm{s}}$ and $D_{\mathrm{m}}$ are the electron and magnon spin diffusion constants. 
The off-diagonal coefficients $D_{\mathrm{sm}}$ and $D_{\mathrm{ms}}$ represent cross-diffusion processes: a gradient of magnon accumulation can induce an electron spin current, while a gradient of electron spin accumulation can drive a magnon spin current. 
These cross terms embody the mutual drag and conversion between electronic and magnonic spin transport.

By combining the coupled continuity relations, Eq.~(\ref{eq:current-continuity}), with the drift–diffusion forms of the electron and magnon spin currents, Eq.~(\ref{eq:drift-diffusion}), one obtains diffusion equations for the nonequilibrium spin densities of electrons and magnons, which can be written compactly as
\begin{equation}
\label{eq:coupled-diffusion}
\begin{pmatrix}
\nabla^2_{\mathbf r}\delta n_{\mathrm{s}}(\mathbf r) \\
\nabla^2_{\mathbf r}\delta n_{\mathrm{m}}(\mathbf r)
\end{pmatrix}
=
\begin{pmatrix}
\lambda_{\mathrm{s}}^{-2} & \lambda_{\mathrm{sm}}^{-2} \\
\lambda_{\mathrm{ms}}^{-2} & \lambda_{\mathrm{m}}^{-2}
\end{pmatrix}
\begin{pmatrix}
\delta n_{\mathrm{s}}(\mathbf r) \\
\delta n_{\mathrm{m}}(\mathbf r)
\end{pmatrix}.
\end{equation}
By expressing the diffusion-coefficient and relaxation-rate matrices as
\begin{equation}\label{eq:D&tau-matrices}
\mathbf{D}=
\begin{pmatrix}
D_{\mathrm{s}} & -D_{\mathrm{sm}}\\ -D_{\mathrm{ms}} & D_{\mathrm{m}}
\end{pmatrix},
\qquad
\boldsymbol{\tau}^{-1}=
\begin{pmatrix}
\tau_{11}^{-1} & \tau_{12}^{-1}\\ \tau_{21}^{-1} & \tau_{22}^{-1}
\end{pmatrix},
\end{equation}
the coefficients in Eq.~(\ref{eq:coupled-diffusion}) follow compactly from
\begin{equation}\label{eq:Lambda-matrix}
\boldsymbol{\Lambda}\equiv
\mathbf{D}^{-1}\,\boldsymbol{\tau}^{-1}=\begin{pmatrix}
\lambda_{\mathrm{s}}^{-2} & \lambda_{\mathrm{sm}}^{-2}\\ \lambda_{\mathrm{ms}}^{-2} & \lambda_{\mathrm{m}}^{-2}
\end{pmatrix}.
\end{equation}
with the entries given by 
\begin{subequations}
\label{eq:lambda_entries}
\begin{align}
\lambda_{\mathrm{s}}^{-2}   &= \frac{ D_{\mathrm{m}}\,\tau_{11}^{-1} + D_{\mathrm{sm}}\,\tau_{21}^{-1} }{\det{(\mathbf{D})}}, 
\label{eq:lambda_s}\\
\lambda_{\mathrm{sm}}^{-2}&= \frac{ D_{\mathrm{m}}\,\tau_{12}^{-1} + D_{\mathrm{sm}}\,\tau_{22}^{-1} }{\det{(\mathbf{D})}},
\label{eq:lambda_sm}\\
\lambda_{\mathrm{ms}}^{-2}&= \frac{ D_{\mathrm{ms}}\,\tau_{11}^{-1} + D_{\mathrm{s}}\,\tau_{21}^{-1} }{\det{(\mathbf{D})}},
\label{eq:lambda_ms}\\
\lambda_{\mathrm{m}}^{-2}   &= \frac{ D_{\mathrm{ms}}\,\tau_{12}^{-1} + D_{\mathrm{s}}\,\tau_{22}^{-1} }{\det{(\mathbf{D})}}.
\label{eq:lambda_m}
\end{align}
\end{subequations}
The coupled drift–diffusion equations were derived by Cheng \emph{et al.} to investigate magnon contributions to the linear magnetoresistance of magnetic bilayers in the current-perpendicular-to-plane (CPP) geometry~\cite{yhCheng17PRB_magnon-el}. Here we rederive these equations in a self-contained form to provide additional insight into electron–magnon cross diffusion and, more importantly, to uncover the role of electrically excited magnons in unidirectional magnetoresistance. 

Several remarks in order. 
(i) In the decoupled limit $D_{\mathrm{sm}}=D_{\mathrm{ms}}=\tau_{12}^{-1}=\tau_{21}^{-1}=0$ and $\tau_{11}\rightarrow \frac{1}{2}\tau_{\uparrow\downarrow}$, $\tau_{22}\rightarrow \tau_{\mathrm{th}}$ (cf.~Appendix~\ref{appendix:D&tau matrices}); hence Eqs.~(\ref{eq:lambda_s}) and (\ref{eq:lambda_m}) reduce to the familiar forms:  $\lambda^0_{\mathrm{s}}=\sqrt{\frac{1}{2}D^0_{\mathrm{s}}\tau_{\uparrow\downarrow}}$~\cite{Fert&Valet93PRB_cpp-GMR} and $\lambda^0_{\mathrm{m}}=\sqrt{D^0_{\mathrm{m}}\tau_{\mathrm{th}}}$~\cite{ZZ12PRB_spin-convertance}, where the superscript ‘0’ denotes bare quantities, i.e., their values in the absence of electron–magnon scattering. Also note that for both electrons and magnons, spin diffusion requires that number-conserving relaxation processes occur much faster than number-nonconserving ones.~\footnote{For electrons, the relaxation time $\tau_{\alpha}$ ($\alpha=\uparrow,\downarrow$) conserves the number of electrons within each spin channel, while the spin-flip relaxation time $\tau_{\uparrow\downarrow}$ changes the relative spin populations; the spin-diffusion regime is valid when $\tau_{\uparrow\downarrow}\gg\tau_{\alpha}$. For magnons, the relaxation time $\tau_\mathrm{m}$ conserves the total magnon number, whereas the thermalization time $\tau_{\mathrm{th}}$ accounts for magnon-number--nonconserving processes; diffusive magnon transport requires $\tau_{\mathrm{th}}\gg\tau_\mathrm{m}$.}
(ii) The interconversion of electron and magnon spin currents (when $\tau_{12}^{-1},\tau_{21}^{-1}\!\neq\!0$) and cross diffusion between them (when $D_{\mathrm{sm}},D_{\mathrm{ms}}\!\neq\!0$) renormalize the effective decay rates via the mixed terms in Eq.~(\ref{eq:lambda_entries}), producing two coupled decay lengths given by the eigenvalues of $\boldsymbol{\Lambda}$.
(iii) Stability requires the positivity of both $\det(\boldsymbol{\Lambda})$ and $\lambda_{\mathrm{sm}}^{-2}\lambda_{\mathrm{ms}}^{-2}$, ensuring $\boldsymbol{\Lambda}$ has positive eigenvalues (real decay lengths). 
(iv) No symmetry is assumed between $D_{\mathrm{sm}}$ and $D_{\mathrm{ms}}$ or between $\tau_{12}^{-1}$ and $\tau_{21}^{-1}$, as the underlying electron and magnon subsystems obey different statistics. Any reciprocity relations, if present, would arise only within a specific microscopic model.

The full integral forms of the entries of the $\mathbf{D}$ and $\boldsymbol{\tau}^{-1}$ matrices---somewhat lengthy and not especially illuminating---are collected in  Appendix~\ref{appendix:D&tau matrices} for completeness. They are intended as a working reference for readers who wish to carry out quantitative analysis of transport properties (including UMR) in different parameter regimes that involve these matrices, complementing the approximate analytical expressions in Ref~\cite{yhCheng17PRB_magnon-el}. In addition, to provide a quantitative sense of the electron–magnon cross-diffusion effects, the calculated elements of the diffusion-coefficient and relaxation-rate matrices at room temperature and zero magnetic field are provided in Table~\ref{table:III-FM|NM} of Appendix~\ref{appendix:tables}. 

\subsection{Out-of-plane linear spin transport in FM$|$NM bilayers}

\begin{figure}[t]
\includegraphics[width=0.45\textwidth]{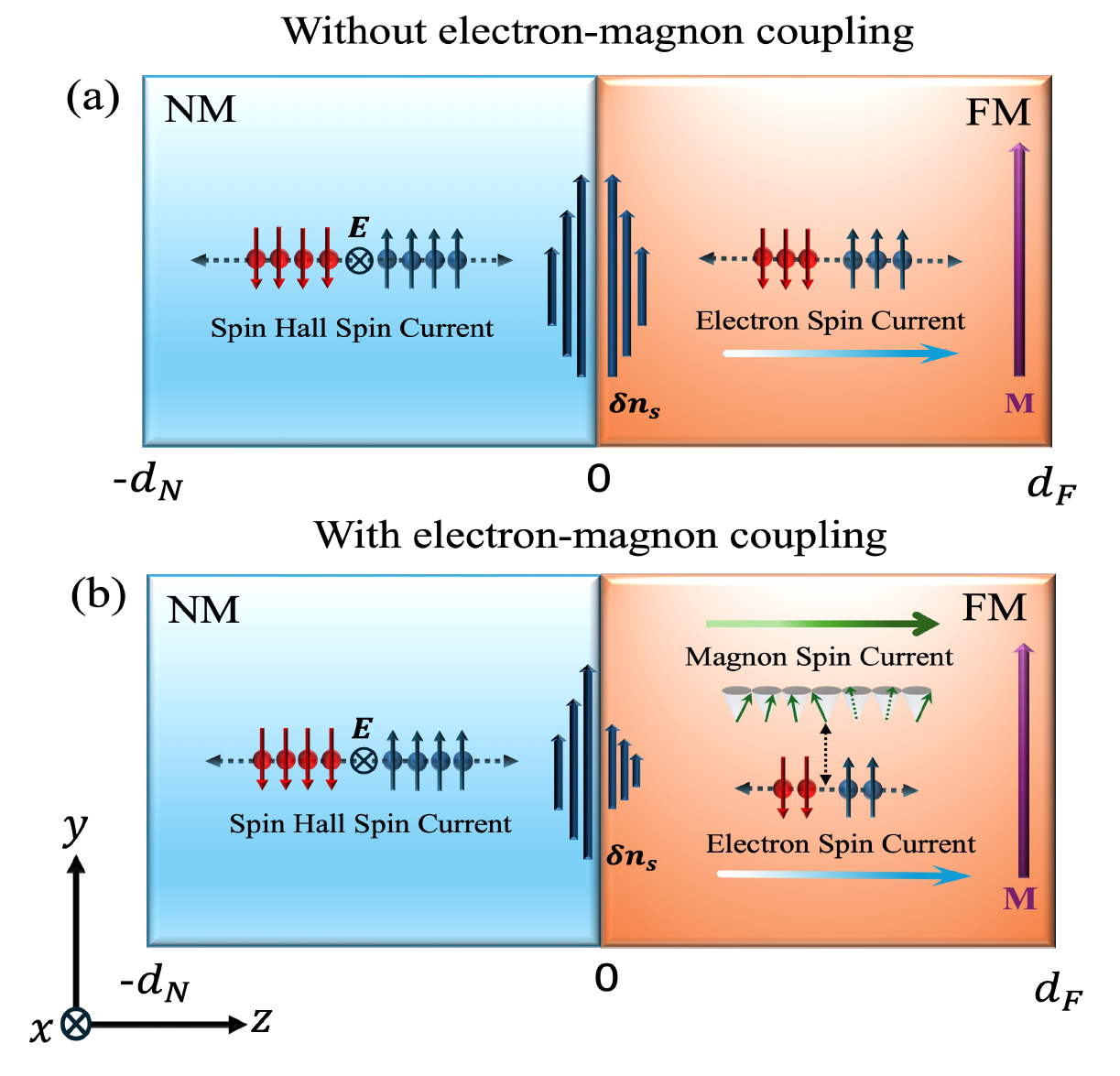}
  \caption{Schematic illustration of spin transport carried by conduction electrons and magnons in an NM$|$FM bilayer. (a) Without electron–magnon interaction: spin-Hall spin current generated in the NM layer is injected into the FM, where both spin accumulation (as represented by dark blue arrows near the interface) and spin current are continuous across the interface, resulting solely in electron spin diffusion in the FM layer. (b) With electron–magnon interaction: the spin current at the NM side of the interface is partially converted into magnon accumulation, producing diffusive magnon spin current in the FM. This leads to a discontinuity in spin accumulation (reduced value at the FM interface) and coexistence of electron and magnon spin currents in the FM layer.}
  \label{fig:bilayer}
\end{figure}

We now apply the general coupled drift--diffusion framework to the geometry of a FM$|$NM bilayer. To induce UMR, an electric field is applied along the $x$-direction, parallel to the layer plane. The coordinate system is shown in Fig.~\ref{fig:bilayer}, where the $z$-axis is taken perpendicular to the bilayer, with the FM layer occupying $z \geq 0$ and the NM layer occupying $z < 0$. 

Within the FM layer, spin transport along the $z$-direction is governed by a generalized Fick’s law for coupled electron and magnon spin currents [see Eq.~(\ref{eq:drift-diffusion})]:  
\begin{equation}
\label{eq:js&jm(z)}
\begin{pmatrix} 
j_{\mathrm{s},z}(z) \\ 
j_{\mathrm{m},z}(z) 
\end{pmatrix}
=-
\begin{pmatrix} 
D_{\mathrm{s}} & -D_{\mathrm{sm}} \\ 
-D_{\mathrm{ms}} & D_{\mathrm{m}} 
\end{pmatrix}
\frac{d}{dz}
\begin{pmatrix} 
\delta n_{\mathrm{s}}(z) \\ 
\delta n_{\mathrm{m}}(z) 
\end{pmatrix},
\end{equation}
where translational invariance in the $x$–$y$ plane is assumed, so that the transport variables depend only on $z$. To further simplify the discussion, we neglect the anomalous Hall effect in the FM layer (whose contribution to UMR has already been analyzed in Ref.~\cite{M&Z22PRB_AH-UMR}), so that in the current-in-plane (CIP) geometry the applied electric field does not directly drive spin transport across the layers.  

The coupled diffusion equations for the nonequilibrium spin densities then take the form  
\begin{equation}
\label{eq:ns(z)}
\frac{d^2}{dz^2}
\begin{pmatrix}
\delta n_{\mathrm{s}}(z) \\
\delta n_{\mathrm{m}}(z)
\end{pmatrix}
=
\begin{pmatrix}
\lambda_{\mathrm{s}}^{-2} & \lambda_{\mathrm{sm}}^{-2} \\
\lambda_{\mathrm{ms}}^{-2} & \lambda_{\mathrm{m}}^{-2}
\end{pmatrix}
\begin{pmatrix}
\delta n_{\mathrm{s}}(z) \\
\delta n_{\mathrm{m}}(z)
\end{pmatrix},
\end{equation}
with $\lambda_{\mathrm{s}}$, $\lambda_{\mathrm{m}}$, $\lambda_{\mathrm{sm}}$, and $\lambda_{\mathrm{ms}}$ the characteristic diffusion lengths of the coupled system.  

The general solutions for the nonequilibrium spin densities in the FM layer are linear combinations of the two eigenmodes,  
\begin{subequations}
\begin{align}\label{eq:gen-sol-ns}
    \delta n_{\mathrm{s}}(z) &= A_{F} e^{z/\lambda_+} + B_{F} e^{-z/\lambda_+} \nonumber\\
    &\quad + C_{F} e^{z/\lambda_-} + D_{F} e^{-z/\lambda_-}, \\  
\label{eq:gen-sol-ms}
    \delta n_{\mathrm{m}}(z) &= \alpha_{+}\!\left(A_{F} e^{z/\lambda_+} + B_{F} e^{-z/\lambda_+}\right) \nonumber \\
    &\quad + \alpha_{-}\!\left(C_{F} e^{z/\lambda_-} + D_{F} e^{-z/\lambda_-}\right),
\end{align}
\end{subequations}
where the two characteristic diffusion lengths $\lambda_{\pm}$ are given by the eigenvalues of the matrix $\boldsymbol{\Lambda}$ in Eq.~(\ref{eq:Lambda-matrix}),  
\begin{equation}\label{eq:lambda_pm}
\lambda_{\pm}^{-2} = \tfrac{1}{2}\left( \lambda_{\mathrm{s}}^{-2} + \lambda_{\mathrm{m}}^{-2}  
\pm  \sqrt{\big( \lambda_{\mathrm{s}}^{-2} - \lambda_{\mathrm{m}}^{-2} \big)^2 + 4\lambda_{\mathrm{sm}}^{-2}\lambda_{\mathrm{ms}}^{-2}} \right).
\end{equation}
In the decoupled limit, $\lambda_{\mathrm{sm}}^{-2} = \lambda_{\mathrm{ms}}^{-2} = 0$, the eigenvalues reduce to the diagonal entries,  
$\ell_{+}^{-2} = \lambda_{\mathrm{s}}^{-2}$ and $\ell_{-}^{-2} = \lambda_{\mathrm{m}}^{-2}$,  
corresponding to independent electron and magnon diffusion channels. The mode–mixing coefficients $\alpha_{\pm}$, defined as the magnon-to-electron weight of each eigenmode, take the equivalent forms
$\alpha_{\pm} = -(\lambda_{\mathrm{s}}^{-2}-\lambda_{\pm}^{-2})/\lambda_{\mathrm{sm}}^{-2} 
= -\lambda_{\mathrm{ms}}^{-2}/(\lambda_{\mathrm{m}}^{-2}-\lambda_{\pm}^{-2})$. 
Their overall normalization is absorbed into the mode amplitudes 
$A_F,B_F,C_F,D_F$ in the general solution.

In the NM layer ($z<0$), the spin drift–diffusion equation governing transport along $z$ is  
\begin{equation}\label{eq:js-NM}
j_{\mathrm{s},z}(z) = -\frac{\sigma_{0,N}}{ \mathcal{N}_{\mathrm{e}}(\epsilon_F)} \frac{d}{dz}\,\delta n_{\mathrm{s}}(z) 
+ \theta_{\mathrm{sh}} \sigma_{0,N} E_x,
\end{equation}
where $\sigma_{0,N}=\nu_N n_{0,N}$ is the Drude conductivity of the NM,  
$\theta_\mathrm{sh}$ is the spin Hall angle, $\mathcal{N}_{\mathrm{e}}^\uparrow(\epsilon_F)=\mathcal{N}_{\mathrm{e}}^\downarrow(\epsilon_F)=\frac{1}{2}\mathcal{N}_{\mathrm{e}}(\epsilon_F)$ 
denotes the spin-resolved density of states at the Fermi level, and $\delta n_{\mathrm{s}}(z)$ is the spin accumulation in NM.  

The second term represents the spin Hall current driven by the in-plane electric field, which serves as the primary source of spin injection into the bilayer. In contrast to the ferromagnetic layer, the NM lacks magnetic order and therefore does not support magnon transport; only electron spin currents contribute here.

The spin accumulation satisfies the diffusion equation
\begin{equation}\label{eq:ns(z)-diffusion}
    \frac{d^2}{dz^2}\,\delta n_{\mathrm{s}}(z) - \frac{\delta n_{\mathrm{s}}(z)}{\lambda_N^2} = 0,
\end{equation}
with $\lambda_N$ the spin diffusion length. Its general solution reads
\begin{equation}\label{eq:gen-sol-ns(z)}
    \delta n_{\mathrm{s}}(z) = A_N e^{z/\lambda_N} + B_N e^{-z/\lambda_N}.
\end{equation}
The coefficients $A_N$ and $B_N$, together with those for the FM layer ($A_F$, $B_F$, $C_F$, $D_F$), constitute six integration constants that are determined by boundary conditions at the FM$|$NM interface and the outer surfaces of the bilayer, thereby fully specifying the spin and magnon accumulations and currents. For the outer surfaces of the FM and NM layers ($z=d_F$ and $z=-d_N$, respectively), open boundaries require both spin and magnon currents to vanish, i.e.,  
\begin{equation}\label{eq:oBCs}
j_{\mathrm{s},z}(d_F)=j_{\mathrm{m},z}(d_F)=0, 
\qquad 
j_{\mathrm{s},z}(-d_N)=0 .
\end{equation} We now turn to the interfacial boundary conditions, where spin angular momentum is transferred between the FM and NM layers and between the electron and magnon subsystems.

\subsection{Interfacial spin angular momentum transfer and interconversion}

Previous works have established the boundary conditions for NM$|$ferromagnetic-insulator (FI) interfaces, where the interfacial electron–magnon interaction mediates angular momentum transfer between conduction electrons in the NM and magnons in the FI~\cite{ZZ12PRB_spin-convertance}. This leads to conservation of the total spin current across the interface, as well as interfacial conversion relations between spin accumulations and currents on either side. In the present case of NM$|$FM bilayers, both conduction electrons and magnons coexist in the ferromagnet, so the boundary conditions must be generalized to capture the richer interconversion processes.  

For the NM$|$FM interface, conservation of total spin angular momentum requires that the spin current injected from the NM be partitioned into both electron and magnon spin currents on the FM side,  
\begin{equation}\label{eq:iBC-j}
    j_{\mathrm{s},z}(0^-) = j_{\mathrm{s},z}(0^+) + j_{\mathrm{m},z}(0^+).
\end{equation}
This condition highlights the essential difference from the NM$|$FI case, where only magnon spin current exists on the ferromagnetic side, so that $j_{\mathrm{s},z}(0^-) = j_{\mathrm{m},z}(0^+)$ directly. In metallic ferromagnets, by contrast, the incoming spin current is distributed between conduction-electron and magnon channels. We also note that, in the earlier work of Cheng \emph{et al.} on FM$|$FM bilayers~\cite{yhCheng17PRB_magnon-el}, continuity of spin current, magnon current, and spin densities was imposed separately, effectively assuming no interfacial exchange coupling. Here, by contrast, the interfacial electron–magnon interaction explicitly mediates angular momentum transfer and interconversion. 

It is also worthy pointing out that in the geometry considered here, the equilibrium in-plane magnetization \(\mathbf{m}\) is perpendicular to the applied electric field and therefore aligned with the current-induced electron spin accumulation \(\boldsymbol{\delta\mu}_s\). In this case, only the longitudinal spin current is relevant, which is unaffected by interfacial spin-mixing effects that act solely on transverse spin currents when \(\mathbf{m}\) and \(\boldsymbol{\delta\mu}_s\) are noncollinear~\cite{Brataas2000, Brataas2001}.

In addition to total conservation, electron–magnon scattering at the interface gives rise to discontinuities in spin current and electron spin accumulation, described by  
\begin{subequations}\label{eq:iBC-Gme-Gem}
\begin{align}\label{eq:iBC-Gme}
    j_{\mathrm{s},z}(0^+) - j_{\mathrm{s},z}(0^-) = G_{\mathrm{me}}\,\delta n_{\mathrm{m}}(0^+), \\ 
    \label{eq:iBC-Gem}
    j_{\mathrm{m},z}(0^+) = G_{\mathrm{em}}\,[\,\delta n_{\mathrm{s}}(0^-) - \delta n_{\mathrm{s}}(0^+)\,],
    \end{align}
\end{subequations}
where $G_{\mathrm{em}}$ and $G_{\mathrm{me}}$ denote the interfacial spin convertances~\cite{ZZ12PRB_spin-convertance}. Physically, $G_{\mathrm{em}}$ characterizes the conversion of electron spin accumulation in the NM into magnon spin current in the FM, while $G_{\mathrm{me}}$ describes the reciprocal process, in which magnon accumulation in the FM generates an electronic spin current.  

In the absence of interfacial electron--magnon scattering, the boundary conditions 
[Eqs.~(\ref{eq:iBC-Gme}) and (\ref{eq:iBC-Gem})] enforce continuity of the electronic spin current and vanishing 
of the interfacial magnon spin current. When interfacial electron–magnon scattering is included and the ferromagnetic layer is insulating (FM$\rightarrow$FI), both the electron spin current and the spin accumulation vanish inside the FI, and the boundary conditions for NM$|$FM bilayers therefore reduce to those established for NM$|$FI bilayer systems~\cite{Takahashi10JPCS_Gem,ZZ12PRB_spin-convertance,Bender12PRL_Gem}. We retain the spin convertances $G_{\mathrm{em}}$ and $G_{\mathrm{me}}$ as those derived for NM$|$FI bilayers, assuming  the NM$|$FM interface is sufficiently rough such 
that electron and magnon momenta are uncorrelated during interfacial electron-magnon scattering---in contrast to bulk electron–magnon scattering where momentum conservation is preserved. Note that interfacial roughness may relax momentum correlations between electrons and magnons, but is not expected to affect the transfer of spin angular momentum normal to the interface; consequently, the continuity of the spin angular-momentum current holds even in the presence of roughness.

We also note that, for the interfacial boundary condition [Eq.~(\ref{eq:iBC-Gem})], the present analysis focuses—at leading order—on the linear response of the interfacial magnon current to the electric-field–induced spin accumulation, proportional to the spin Hall angle $\theta_{\mathrm{sh}}$. The resulting UMR coefficient therefore scales linearly with $\theta_{\mathrm{sh}}$. Extending the boundary condition to include quadratic response to the spin accumulation would give rise to a UMR contribution cubic in $\theta_\mathrm{sh}$~\cite{Duine19PRB_magnon-USMR}.

\subsection{In-plane nonlinear charge transport and UMR coefficient}
\label{subsec:IP-UMR}
To examine the in-plane nonlinear charge transport, we begin with the expression for the charge current density in terms of spin-resolved carrier densities and mobilities. Reorganizing this relation makes explicit how spin accumulation couples to spin-dependent mobility, which is the microscopic origin of UMR in the FM layer. We begin with the general expression for the in-plane charge current density as the sum over the two spin channels, 
\begin{equation}\label{eq:jcx-sum}
j_{c,x}(z)=E_x\sum_{\sigma=\uparrow,\downarrow}\nu_\sigma\,n_\sigma(z),
\end{equation}
which can be rewritten as   
\begin{equation}\label{eq:jcx-dns0}
j_{c,x}(z)=\bar{\nu}\big[n_{+}(z)+P_{\nu}\,n_{-}(z)\big]E_x,
\end{equation}
wherein $n_{\pm}(z)\equiv n_\uparrow(z)\pm n_\downarrow(z)$, $\bar{\nu}\equiv (\nu_\uparrow +\nu_\downarrow)/2$ is the spin-averaged electron mobility, and $P_{\nu}\left(\equiv \frac{\nu_\uparrow -\nu_\downarrow}{\nu_\uparrow +\nu_\downarrow}\right)$ is the spin asymmetry of mobility. Note that 
$n_{-}(z)$ coincides with the spin accumulation $\delta n_\mathrm{s}(z)$ and $n_{+}(z)\equiv n_\uparrow(z)+n_\downarrow(z)$ is the total (local) density of conduction electrons. In the absence of local charge accumulation, $n_\uparrow(z)+n_\downarrow(z)=n_0$, where $n_0$ is the total equilibrium electron density; in this case, only the spin accumulation $\delta n_\mathrm{s}(z)$ can be electrically induced and vary spatially. It follows that   
\begin{equation}\label{eq:jcx-dns1}
\,j_{c,x}(z)=\sigma_{0,F} E_x+P_{\nu}\bar{\nu}\,\delta n_\mathrm{s}(z)\,E_x\, ,
\end{equation}
with $\sigma_{0,F}=\bar{\nu} n_0$ is the (linear) Drude conductivity of the FM. Equation~(\ref{eq:jcx-dns1}) shows that microscopically, UMR arises from the combined effects of spin-asymmetry of mobility ($P_{\nu}\neq 0$) and electrically induced spin accumulation ($\delta n_\mathrm{s} \neq 0$).

In the NM layer, where $P_{\nu}=0$, the UMR effect vanishes even with a net spin accumulation.  In the FM layer, however, the coexistence of spin-dependent mobility and spin accumulation gives rise to nonlinear charge transport under the applied electric field, yielding a finite UMR contribution.

To quantify the UMR effect, we introduce the UMR coefficient defined as
\begin{equation}
\label{eq:cUMR}
\zeta_{\text{UMR}} \equiv 
\frac{\bar{\sigma}(E_x) - \bar{\sigma}(-E_x)}{\sigma_0 E_x},
\end{equation}
where $\bar{\sigma}(E_x)$ denotes the spatially averaged conductivity of the bilayer under applied field $E_x$, and $\sigma_0$ is the linear (Drude) conductivity of the bilayer. This definition is independent of the external electric field, reflecting intrinsic properties of the bilayer. Moreover, since $\zeta_{\text{UMR}}$ has the dimension of the inverse electric field, it carries a clear physical meaning: it represents the characteristic field scale at which the nonlinear conductivity becomes comparable to its linear counterpart.

\section{Results and Discussion}
\label{sec:R&D}

In this section, we compute the UMR coefficient for a representative FM$|$NM bilayer using realistic material parameters and analyze its dependence on strength of exchange coupling, external magnetic field, layer thickness, and temperature, thereby identifying the role of magnons in UMR and experimentally testable signatures.

\subsection{Roles of electron–magnon interactions}

Magnons contribute to UMR through the transfer of angular momentum and momentum between electrons and magnons. These exchanges are enabled by electron–magnon interactions, which couple the two subsystems at interfaces and within the bulk of the FM. This physical picture is illustrated in Fig.~\ref{fig:bilayer}, which contrasts spin  transport without and with electron-magnon interactions. In the absence of electron--magnon coupling, the spin-Hall spin current from the NM flows into the FM entirely via electron spin transport. With electron-magnon interactions, this spin current is partially converted into magnon currents and accumulations, giving rise to coexisting electron and magnon diffusion inside the FM. This conversion simultaneously reduces the electronic spin accumulation in the FM layer near the interface and, consequently, the UMR, since the latter scales with the net spin accumulation therein [see Eq.~(\ref{eq:jcx-dns1})].

\begin{figure}[t]
\includegraphics[width=0.45\textwidth]{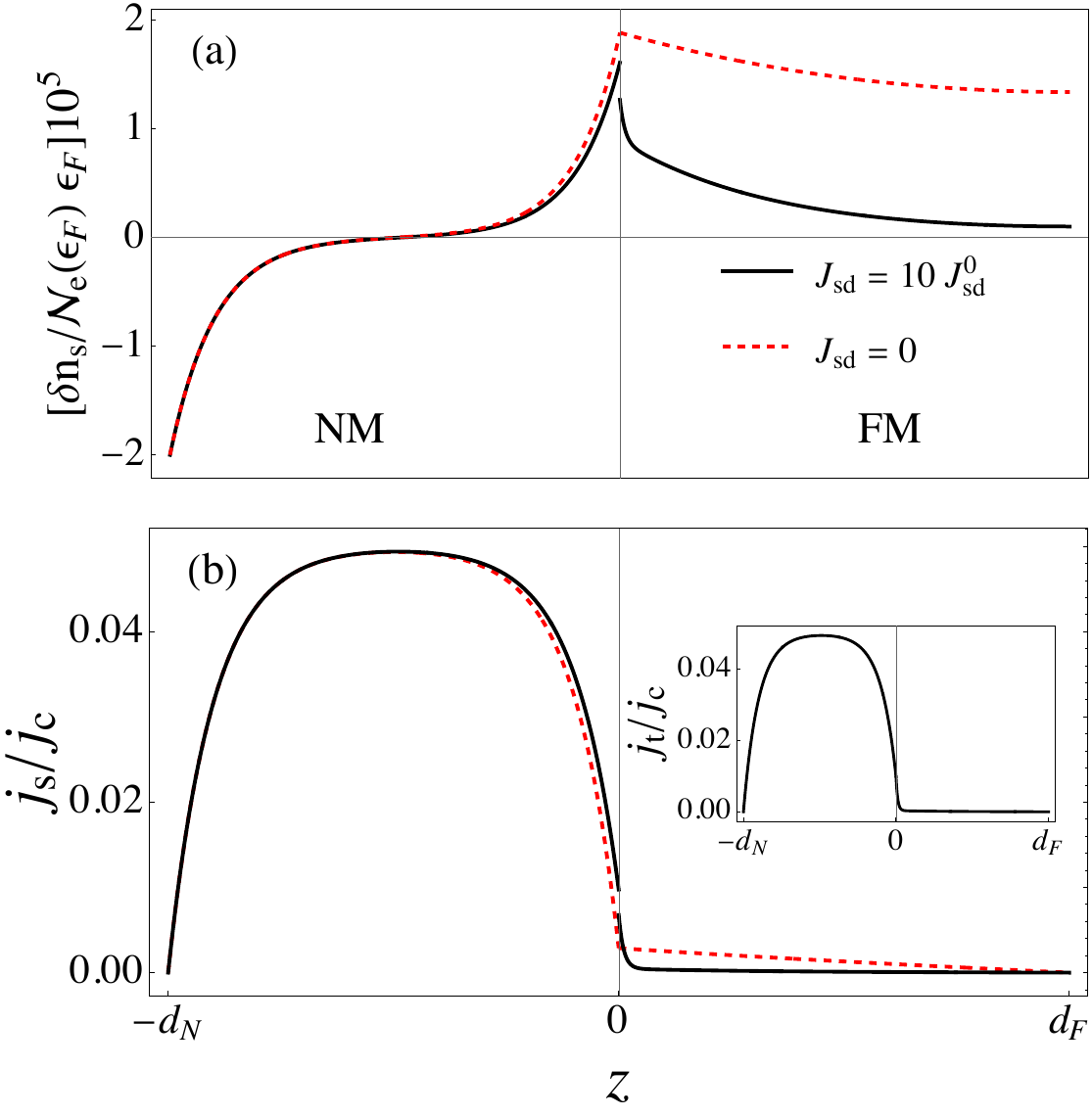}
  \caption{Spatial profiles of spin accumulation and spin current in an FM$|$NM bilayer, with and without electron–magnon interactions. (a) Spin accumulation $\delta n_{\mathrm{s}}(z)$: for $J_{\mathrm{sd}}=0$ the profile is continuous across the interface (dashed red curve), while for $J_{\mathrm{sd}}\neq 0$ a discontinuity appears at the boundary and the spin accumulation inside the FM is reduced (solid black curve). (b) Electron spin current density $j_{\mathrm{s},z}(z)$: without electron--magnon interactions, the current is continuous across the interface at $z=0$ (dashed red curve); with electron--magnon interactions, $j_{\mathrm{s},z}$ shows a discontinuity at $z=0$ and is reduced in the FM (solid black curve), reflecting conversion into magnon currents. Inset shows the continuity of total spin angular momentum current across the interface when $J_{\mathrm{sd}} \neq 0$. The calculations were performed using $J_{\mathrm{sd}}^0 = 0.1\,\text{eV}$, $E_x=10^{-4}$ V/nm, and fixed layer thicknesses $d_N=d_F=50$ nm.}
  \label{fig:ns&js-z} 
\end{figure}

To better illustrate the physical picture described above, we examine the spatial profiles of spin accumulation and spin current density across the bilayer. For vanishing exchange coupling ($J_\mathrm{sd}=0$), the electron spin accumulation $\delta n_\mathrm{s}$ is continuous across the interface and decays smoothly inside the FM. For finite coupling ($J_\mathrm{sd}>0$), $\delta n_\mathrm{s}$ instead exhibits a discontinuity at the interface and is reduced throughout the FM layer. A similar feature appears in the spin current profile, Fig.~\ref{fig:ns&js-z}(b), which develops a discontinuity at the interface and decreases inside the FM due to partial conversion into magnon currents, with the total spin angular momentum remaining continuous across the interface (see inset of Fig.~\ref{fig:ns&js-z}). These behaviors result from the angular momentum transfer between electron and magnon subsystems, encoded at the interface through the boundary conditions [Eqs.~(\ref{eq:iBC-j}) and (\ref{eq:iBC-Gme-Gem})] and within the FM through cross diffusion [Eq.~(\ref{eq:js&jm(z)})].

\begin{figure}[t] 
  \includegraphics[width=0.45\textwidth]{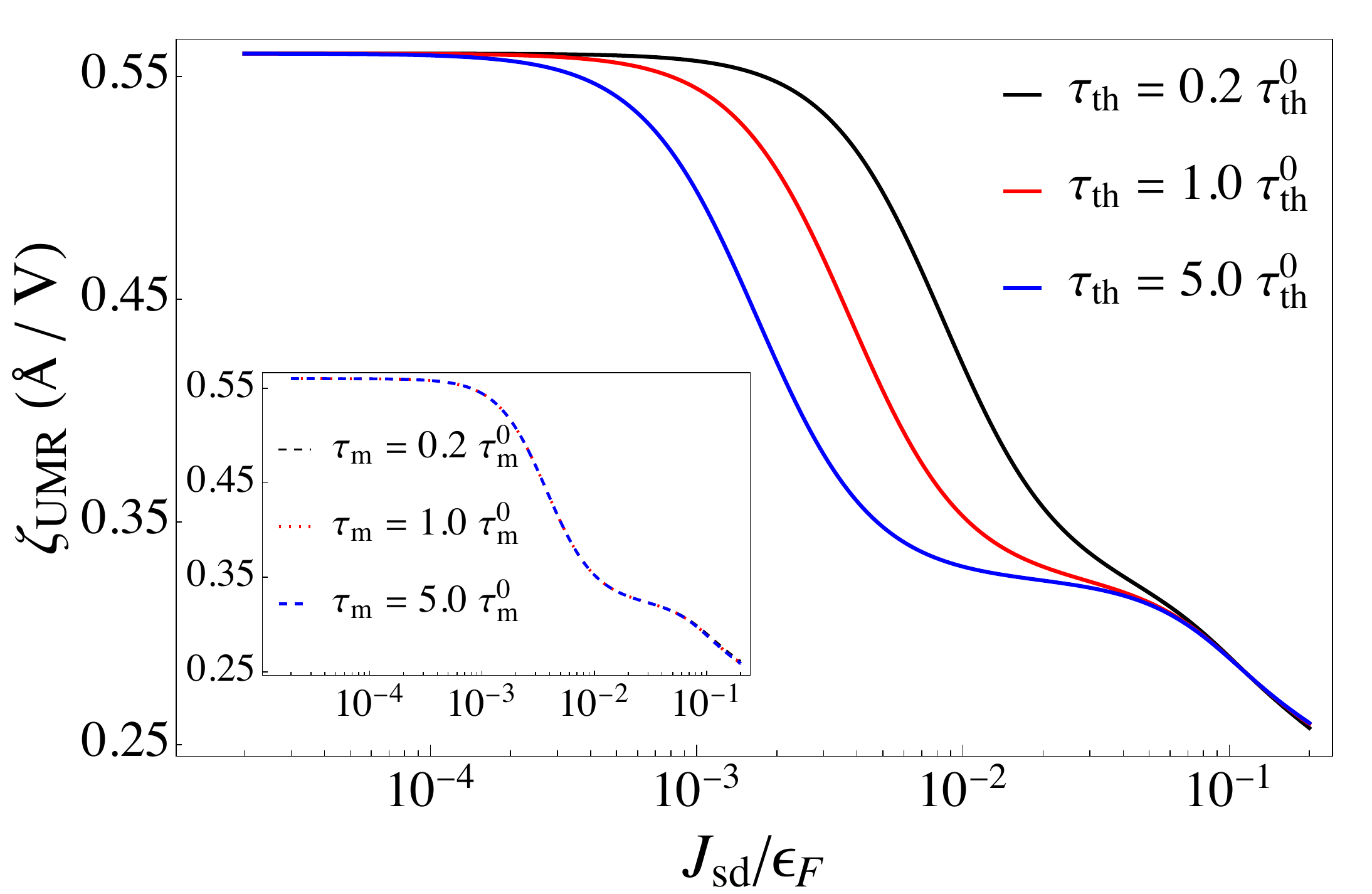}
  \caption{UMR coefficient $\zeta_{\mathrm{UMR}}$ as a function of exchange coupling. 
  The solid curves show results for different magnon thermal relaxation times $\tau_{\mathrm{th}}$ ($\tau^0_{\mathrm{th}}=100\, \text{ps}$). 
  Inset: dependence of $\zeta_{\mathrm{UMR}}$ on exchange coupling for different magnon momentum-relaxation times $\tau_{\mathrm{m}}$ ($\tau^0_{\mathrm{m}}=10\, \text{ps}$).}
  \label{fig:UMR-vs-Jsd} 
\end{figure}

Having established the microscopic mechanisms through spatial profiles of spin accumulation and currents, we now turn to the macroscopic observable—the UMR coefficient. Figure~\ref{fig:UMR-vs-Jsd} shows its dependence on the strength of the exchange interaction  $J_\mathrm{sd}$~\footnote{For simplicity, we take the interfacial and bulk exchange couplings to be characterized by the same parameter $J_\mathrm{sd}$. In principle, they may differ in magnitude because the local electronic environment and hybridization at the interface can modify the $s$–$d$ exchange strength.} for several values of the magnon thermal relaxation time $\tau_{\mathrm{th}}$.  Overall, the UMR decreases with increasing $J_\mathrm{sd}$: stronger electron--magnon interaction transfers more spin angular momentum from the electronic subsystem into magnons, thereby reducing the spin accumulation available to contribute to UMR. In the intermediate $J_\mathrm{sd}$ regime, the UMR is smaller for larger $\tau_{\mathrm{th}}$, since faster magnon thermalization (smaller $\tau_{\mathrm{th}}$) allows more electron spin accumulation to contribute to UMR. The curves converge in both limits: In the weak-$J_\mathrm{sd}$ limit, electron and magnon spin transport are only weakly coupled, and the UMR approaches a constant value set by spin-dependent elastic scattering alone~\cite{Zhang&Vignale16PRB_USMR}; in the strong-$J_\mathrm{sd}$ limit, the UMR becomes independent of $\tau_{\mathrm{th}}$. This reflects saturated angular mometnum conversion between electron and magnon subsystems: the electronic spin accumulation in the FM is maximally suppressed, so the details of magnon thermalization no longer affect the net FM spin accumulation or the resulting UMR coefficient. The inset shows the UMR as a function of $J_\mathrm{sd}$ for several values of the magnon momentum-relaxation time $\tau_\mathrm{m}$: the same $J_\mathrm{sd}$ dependence is observed, but the curves are insensitive to $\tau_\mathrm{m}$, consistent with the fact that $\tau_\mathrm{m}$ governs processes conserving magnon number.

\subsection{Variation of UMR with magnetic field and magnon gap}

\begin{figure}[t] \includegraphics[width=0.45\textwidth]{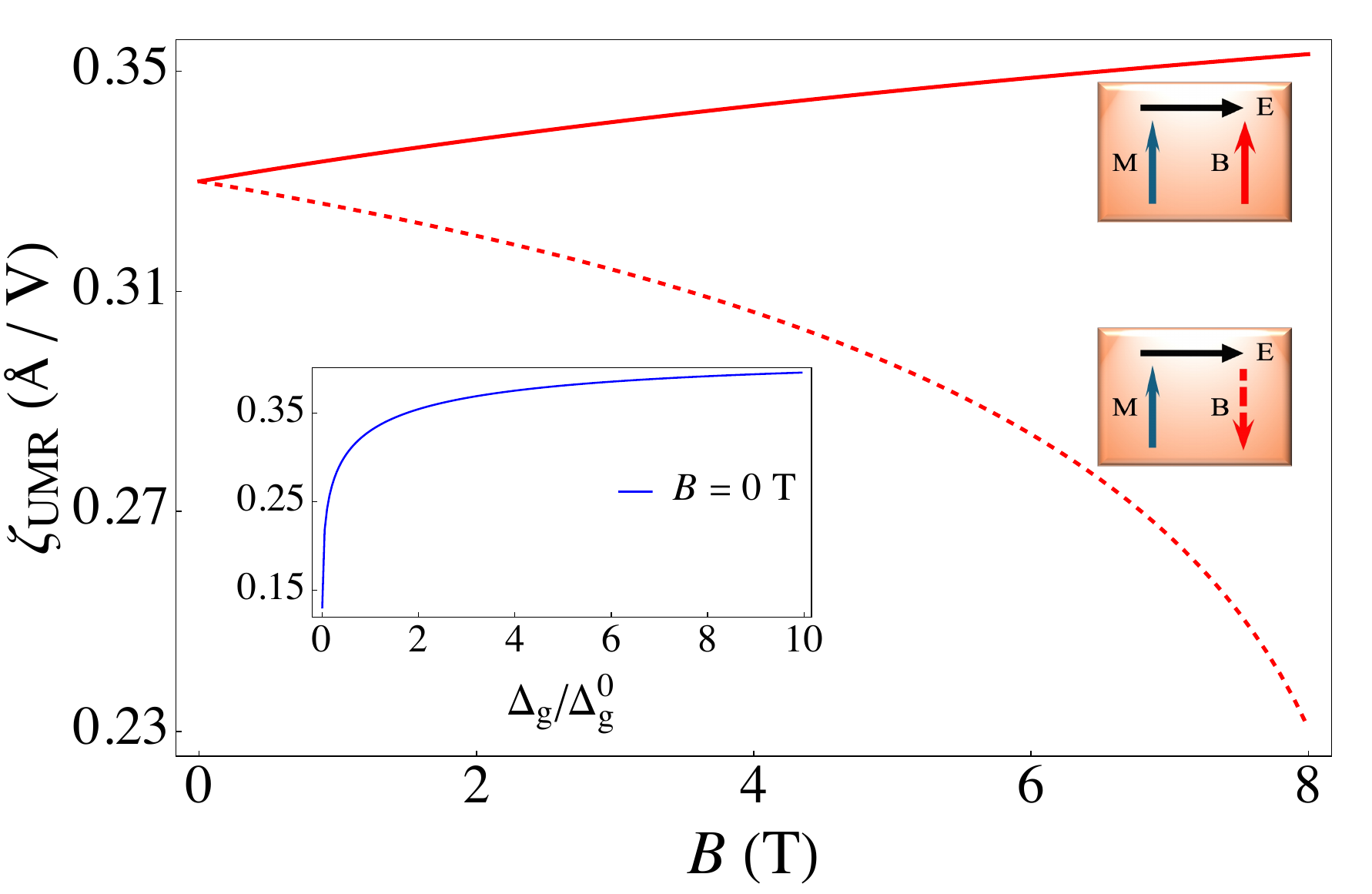}
  \caption{UMR coefficient $\zeta_{\mathrm{UMR}}$ as a function of external magnetic field $B$. The solid red curve corresponds to $B$ applied parallel to the magnetization, and the dashed red curve to $B$ applied antiparallel. Inset: $\zeta_{\mathrm{UMR}}$ as a function of intrinsic magnon gap $\Delta_\mathrm{g}$ (scaled with $\Delta^0_\mathrm{g}=1 \, \text{meV}$).  }
  \label{fig:UMR-v-B}
\end{figure}

We next examine how the UMR coefficient varies with the external magnetic field and with the intrinsic magnon gap. Figure~\ref{fig:UMR-v-B} shows the UMR coefficient as a function of magnetic field applied parallel or antiparallel to the magnetization. For a parallel field, $\zeta_{\mathrm{UMR}}$ increases with $B$ and eventually saturates at a value determined by the rate of spin angular momentum transfer between the electron and magnon subsystems. This behavior reflects the field-induced increase of the effective magnon excitation energy~\footnote{We neglect dipolar interactions in the magnon dispersion because they affect only the long-wavelength (small-\(q\)) part of the spin-wave spectrum~\cite{Holstein40PR_HP-transformations}, whereas at larger wave vectors the quadratic exchange term \(\propto q^{2}\) dominates and the dispersion becomes effectively exchange-like. This is illustrated explicitly in Appendix~\ref{appendix:Dipolar-interaction}, where we compare the exchange-only and full (exchange + dipolar) magnon dispersions and show that, over the wave-vector range relevant for transport, they essentially coincide; moreover, we demonstrate that the UMR is dominated by short-wavelength exchange magnons, while long-wavelength magnons contribute weakly due to predominantly forward scattering.
},
\begin{equation}\label{Eq:magnon-disp}
\omega_{q} = A_{\mathrm{ms}} q^{2} + g \mu_B\, \mathbf{m}\!\cdot\!\mathbf{B} + \Delta_\mathrm{g},
\end{equation}
where $A_{\mathrm{ms}}$ is the magnon stiffness, $\mu_B$ the Bohr magneton, $g$ the Land\'e $g$-factor, and $\mathbf{m}$ the unit vector along the magnetization direction, which makes the creation of nonequilibrium magnons more energetically costly. As fewer magnons are excited, less spin angular momentum is transferred from the electronic subsystem into magnons, leaving more interfacial spin accumulation to contribute to UMR. In contrast, for an antiparallel field, the effective gap is reduced by the opposite Zeeman term. The resulting suppression of the magnon gap enhances magnon excitation, increasing angular momentum transfer into magnons and thereby reducing UMR. When the Zeeman contribution nearly cancels the intrinsic gap $\Delta_\mathrm{g}$, the effective total gap approaches zero, leading to a rapid drop of $\zeta_{\mathrm{UMR}}$. The inset of Fig.~\ref{fig:UMR-v-B} shows the explicit dependence on $\Delta_\mathrm{g}$, highlighting that a larger intrinsic gap monotonically enhances UMR until saturation is reached for a given coupling strength $J_\mathrm{sd}$.

Recall that the generation of UMR in metallic bilayers requires two ingredients: (i) the current-induced spin accumulation in the FM layer, and (ii) the spin asymmetry of mobility $P_{\nu}$, such that  $\zeta_{\text{UMR}}\propto P_{\nu}\,\delta n_\mathrm{s}$. In this study, we focus on how nonequilibrium magnons reduce the spin accumulation $\delta n_\mathrm{s}$, while $P_{\nu}$—set by band structure or impurity scattering—remains electric-field independent. This corresponds to the \textit{passive} role of magnons, where they drain spin angular momentum from conduction electrons without altering the mobility spin asymmetry $P_{\nu}$. In this regime, a field antiparallel to the magnetization enhances magnon excitation and decreases UMR, while a parallel field suppresses magnons and increases UMR, as shown in Fig.~\ref{fig:UMR-v-B}.

In principle, however, magnons can also play an \emph{active} role (constructive for UMR) by modifying $P_{\nu}$ itself through electron--magnon scattering. When the spin accumulation is antiparallel to the magnetization, enhanced magnon excitation increases electron scattering and thereby the mobility asymmetry; when the spin accumulation is parallel to the magnetization, suppressed magnon excitation weakens scattering and correspondingly lowers the mobility asymmetry. This mechanism introduces an additional electric-field--induced contribution to $P_{\nu}$ (i.e., $P_{\nu}\propto E_x$) which, together with the equilibrium spin polarization $(n_\uparrow^{0} - n_\downarrow^{0})$, can contribute to UMR. The resulting magnetic-field dependence would be opposite to that in Fig.~\ref{fig:UMR-v-B} and qualitatively resembles trends reported in several bilayer experiments~\cite{Ferguson16APL_UMR-STO(exp),avci2018prl,Gambardelaa25PRL_UMR-magnon(exp)}. A full formulation and careful analysis of this active magnon contribution are needed to assess its relevance to experiment; this lies beyond the scope of the present study and will be pursued elsewhere.

\subsection{Magnon contributions revealed by thickness and temperature dependence}

\begin{figure}[t]  \includegraphics[width=0.45\textwidth]{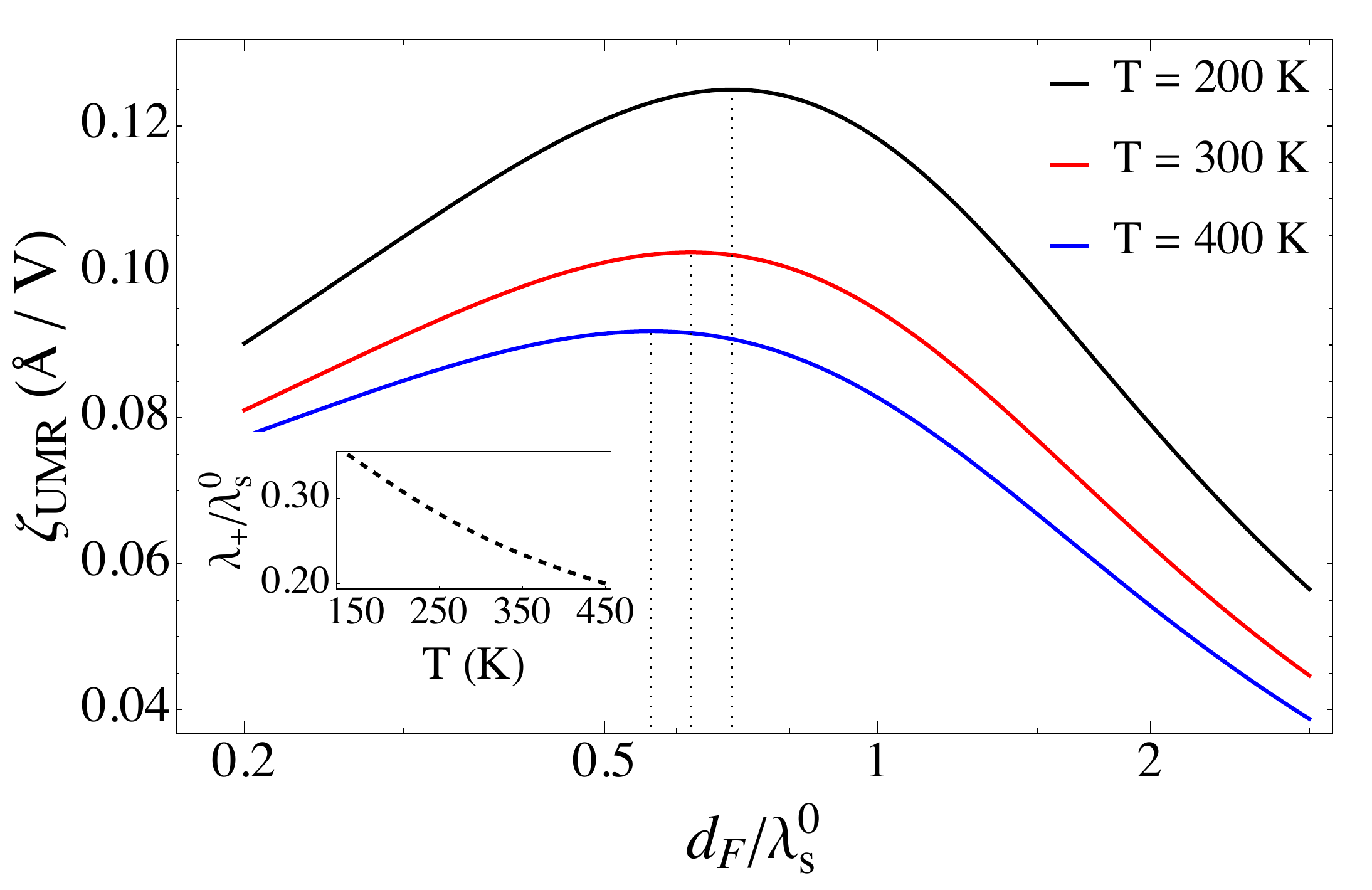}
  \caption{UMR coefficient $\zeta_{\mathrm{UMR}}$ as a function of FM-layer thickness $d_F/\lambda_{\mathrm{s}}^0$ at different temperatures.  
  Inset: normalized characteristic electron diffusion length $\lambda_{+}/\lambda_{\mathrm{s}}^0$ versus temperature for a magnon gap $\Delta_\mathrm{g} = 10^{-3}\Delta_\mathrm{g}^0$.  
  Here $\lambda_{\mathrm{s}}^0$ denotes the bare electron diffusion length in FM. The NM-layer thickness is fixed at 50 nm.\label{fig:UMR-v-thick}}
\end{figure}

Lastly, we investigate how magnon contributions to UMR manifest through their dependence on FM thickness and temperature. Figure~\ref{fig:UMR-v-thick} shows the UMR coefficient $\zeta_{\mathrm{UMR}}$ as a function of FM thickness for several temperatures, with the inset displaying the characteristic electron diffusion length $\lambda_{+}$ as a function of temperature. For a fixed temperature, $\zeta_{\mathrm{UMR}}$ exhibits a peak at a characteristic thickness set by $\lambda_{+}$. At small thicknesses, spin accumulation cannot fully build up near the interface, leading to a reduced UMR. As the thickness increases, the accumulation saturates and UMR reaches a maximum. Beyond the peak, further increase in thickness dilutes the nonequilibrium spin density throughout the FM layer, resulting in a gradual reduction of UMR.

The peak position shifts systematically with temperature: as shown in the inset, $\lambda_{+}$ decreases with increasing $T$ due to enhanced scattering from thermal magnons. As a result, the peak in $\zeta_{\mathrm{UMR}}$ moves to smaller FM thicknesses, reflecting the shortened diffusion length. In addition to this shift, the overall magnitude of the UMR decreases at elevated temperatures, consistent with an increased population of nonequilibrium magnons that absorb spin angular momentum from conduction electrons, reducing the spin accumulation in the FM layer. This trend is also reflected in the inset of Fig.~\ref{fig:UMR-v-thick}, where the effective electron diffusion length $\lambda_{+}$ at higher temperatures deviates more strongly from its value in the decoupled case $\lambda_{\mathrm{s}}^0$. These thickness and temperature dependencies provide clear signatures of magnon involvement in UMR that may be verified experimentally. 

For clarity and to avoid redundancy, we collect the material parameters of the FM$|$NM bilayer in Appendix~\ref{appendix:tables}, wherein we also show the relations among interdependent quantities (cf.~Tables~\ref{table:I-FM} and~\ref{table:II-NM}) and provide order-of-magnitude estimates of key parameters under experimentally accessible conditions (cf.~Table~\ref{table:III-FM|NM}).

\section{Conclusions and outlook}
\label{sec:Conclusion}

We developed a theoretical framework for nonlinear charge transport in metallic NM$|$FM bilayers by extending the coupled electron–magnon transport formalism beyond the linear-response regime and incorporating generalized interfacial boundary conditions that account for electron--magnon scattering at the interface. This framework captures the mutual transfer of spin angular momentum between the electronic and magnonic subsystems and provides a unified description of spin accumulation and spin-current densities across the bilayer, enabling quantitative evaluation of how electron–magnon interactions modify both bulk and interfacial transport.

Within this framework, we showed that (exchange) magnons can play a \emph{passive} role in
UMR generation: by transferring spin angular momentum from conduction electrons
into the magnon subsystem, they reduce the electronic spin accumulation at the
FM$|$NM interface and thereby suppress UMR. The resulting decrease of UMR with
increasing exchange coupling $J_\mathrm{sd}$, the saturation behavior at strong
coupling, and the opposite magnetic-field dependences under parallel versus
antiparallel orientations all consistently support this physical mechanism.  

We further analyzed the impact of electron–magnon scattering and cross diffusion on the thickness and temperature dependences of UMR, finding that the effect peaks at a characteristic FM thickness set by the ‘dressed’ electron spin diffusion length. With increasing temperature, the overall magnitude of the UMR decreases and the peak position shifts toward smaller FM thicknesses, reflecting enhanced thermal magnon scattering. These features provide clear signatures that can be tested experimentally to identify magnonic contributions to UMR. 

Looking ahead, it would be of interest to extend the present framework to include the influence of magnons on both interfacial and bulk spin-dependent momentum relaxation of electrons, and the resulting impact on UMR. In particular, current-induced magnons can modify the momentum relaxation times of conduction electrons in the FM layer—specifically, $\tau_{\uparrow}$ and $\tau_{\downarrow}$ in Eq.~(\ref{eq:KE-electron})—which, in the present study, were treated as fixed phenomenological parameters. Such renormalization would directly affect the spin-dependent mobilities and could, in principle, allow magnons to enhance rather than suppress the UMR. 

Magnons may also influence interfacial spin-dependent scattering, where their interplay with spin accumulation in the adjacent NM layer could lead to additional UMR even in NM$|$FI bilayers. Extending the present theoretical framework to such heterostructures---where UMR has also been observed experimentally~\cite{dWu21PRL_magnon-USMR}---could help address limitations of existing theoretical descriptions~\cite{gGuo18PRB_UMR-FI-NM(theo),Duine19PRB_magnon-USMR} and provide a unified interpretation of UMR, particularly regarding its current, thickness, and spin–Hall–angle dependencies. 

Moreover, in systems hosting spin–momentum–locked surface states, momentum correlations in interfacial electron–magnon scattering may become crucial~\cite{kYasuda16PRL_UMR-TI}, necessitating corresponding modifications to the interfacial boundary conditions [Eqs.~(\ref{eq:iBC-Gme}) and (\ref{eq:iBC-Gem})]. Specifically, the presence of such surface states would lead to modified expressions for \(G_{\mathrm{em}}\) and \(G_{\mathrm{me}}\) that explicitly incorporate momentum conservation—and the associated spin transfer arising from momentum exchange—at the interface. In addition, spin–momentum–locked surface states may contribute directly to nonlinear in-plane charge transport, independent of perpendicular spin or charge transport~\cite{Pan&Zhang18NatPhys_BMER,aDyrdal20PRL_BMR-TI,mm&SZ24PRB_nl-transport-TI}. Exploring these effects more comprehensively—particularly their interplay with magnon-mediated interfacial transport—could yield deeper insights into the underlying magnon-mediated interfacial spin- and momentum-transfer mechanisms and guide future experimental investigations.

\section*{Acknowledgments}
We thank W. Lambrecht, M. Mehraeen, and Yihong Cheng for helpful discussions. 
The work on the electrical generation of magnon current in linear response was partly supported by National Science Foundations.

\appendix
\counterwithin{figure}{section}

\section{Parameters for electron–magnon transport in NM$|$FM bilayers}
\label{appendix:tables}

In this Appendix, we present three tables, i.e., Tables~\ref{table:I-FM}–\ref{table:III-FM|NM}, that compile the numerical values, definitions, and interrelations of parameters used in modeling electron and magnon transport across the NM$|$FM bilayer and in the calculation of the UMR coefficient. These tables serve as a unified reference for implementing the transport matrices introduced in the main text. In addition, Table~\ref{table:III-FM|NM} collects representative values of the diffusion-coefficient and relaxation-rate matrices at room temperature and zero magnetic field, providing quantitative context for assessing the strength of electron–magnon cross-diffusion effects.

\begin{figure}[htbp]  \includegraphics[width=0.45\textwidth]{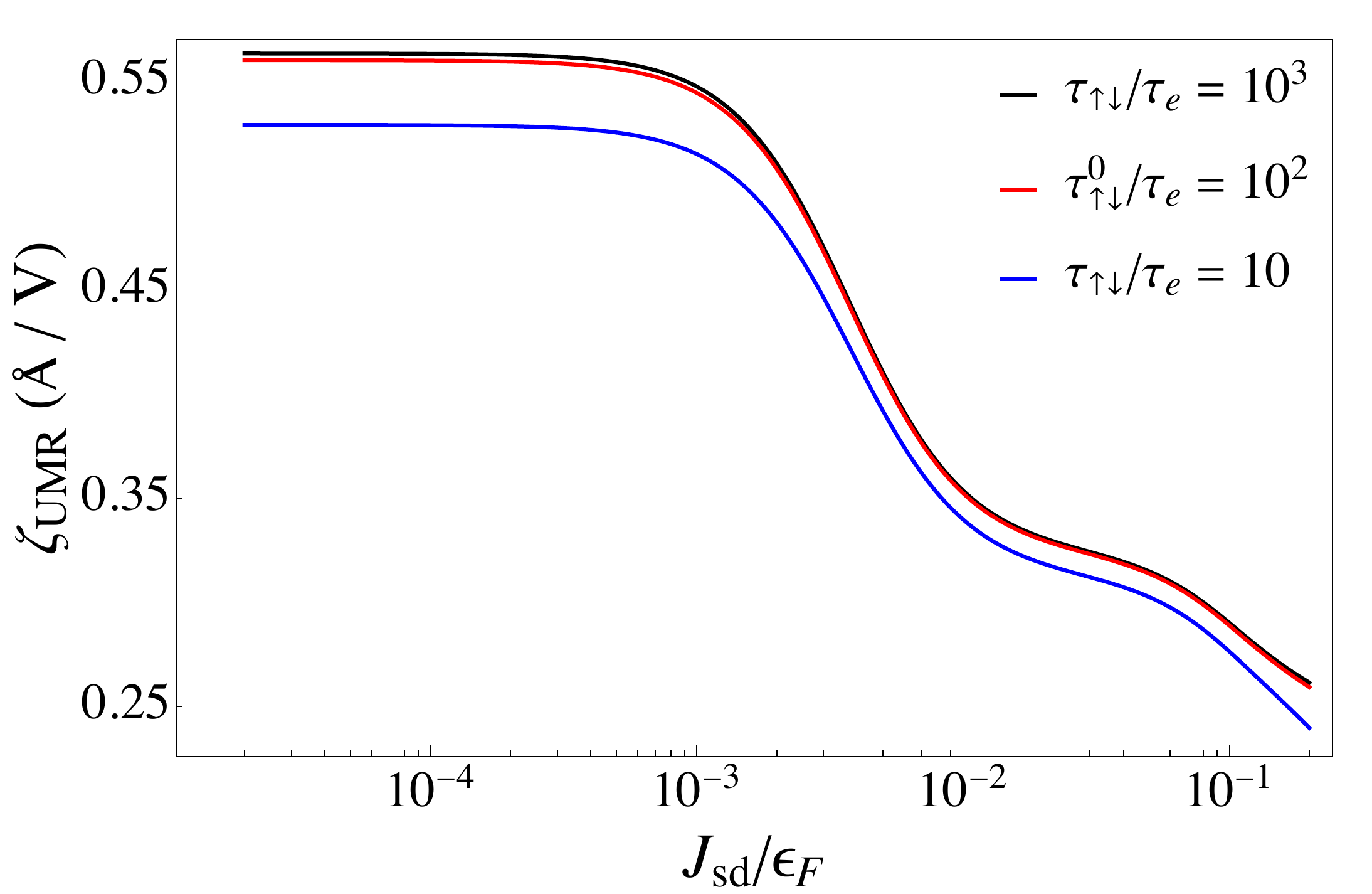}
  \caption{UMR coefficient $\zeta_{\mathrm{UMR}}$ as a function of exchange coupling. 
  The solid curves show results for different electron spin-flip relaxation times $\tau_{\uparrow \downarrow}$ ($\tau^0_{\uparrow \downarrow}=1\, \text{ps}$) with fixed electron momentum relaxation time $\tau_{\mathrm{e}}=0.01$ \text{ps}.~\label{fig:UMR-vs-tud}}
\end{figure}

As the UMR is governed by several key relaxation and scattering parameters that may vary across different materials, we further assess the sensitivity of the UMR coefficient to these parameters. The relaxation processes central to the magnon-mediated UMR include the electron relaxation times \(\tau_{\uparrow}\) and \(\tau_{\downarrow}\), the electron spin-flip relaxation time \(\tau_{\uparrow\downarrow}\), and the magnon relaxation times \(\tau_{\mathrm{m}}\) and \(\tau_{\mathrm{th}}\). In the main text, we showed that the UMR depends on the spin asymmetry of the electron relaxation times (neglecting the spin dependence of the Fermi energy and effective mass), as discussed in Sec.~\ref{subsec:IP-UMR}, and illustrated its dependence on \(\tau_{\mathrm{m}}\) and \(\tau_{\mathrm{th}}\) in Fig.~\ref{fig:UMR-vs-Jsd} and its inset. Here we additionally examine the sensitivity of the UMR coefficient to the spin-flip relaxation time \(\tau_{\uparrow\downarrow}\), as shown in Fig.~\ref{fig:UMR-vs-tud}. We find that the UMR coefficient is relatively insensitive to spin-flip relaxation provided that \(\tau_{\uparrow\downarrow}\) remains longer than the spin-conserving relaxation times: varying \(\tau_{\uparrow\downarrow}\) by two orders of magnitude results in less than a 10\% change in the UMR coefficient.

\begin{table*}
    \caption{Parameters for uncoupled electron and magnon transport in the FM layer}
    \label{table:I-FM}
    \renewcommand{\arraystretch}{1.5}
    \begin{ruledtabular}
    \begin{tabular}{lccc}
        \textbf{Description} & \textbf{Symbol} & \textbf{Expression} & \textbf{Value}  \\ \hline
        Fermi energy & $\epsilon_F$ & -- & $5$ eV \\
         Fermi velocity & $v_F$ & $\sqrt{2\epsilon_F/m_\mathrm{e}^{\ast}}$ & $10^6$ m/s  \\Mean electron momentum relaxation time & $\overline{\tau_\mathrm{e}}$ & $2\tau_\uparrow\tau_\downarrow/(\tau_\uparrow+\tau_\downarrow)$ & $0.02$ ps \\
        Spin asymmetry of electron momentum relaxation time & $p_\tau$ & $(\tau^{\uparrow}-\tau^{\downarrow})/(\tau^{\uparrow}+\tau^{\downarrow})$ & $0.7$  \\
        Electron spin-flip relaxation time & $\tau_{\uparrow \downarrow}$ & -- & $1$ ps~\cite{Cinchetti2006}  \\
        Drude conductivity & $\sigma_{0,F}$ & $e^2 D^0_{\mathrm{s}} \mathcal{N}_\mathrm{e}(\epsilon_F)/(1-p^2_\tau)$~\footnote{The conductivity follows from Mott’s two-current model and the Einstein relation applied to each spin channel, such that the total conductivity is 
$\sigma = \sigma_{\uparrow} + \sigma_{\downarrow}$ 
with 
$\sigma_{\alpha} = e^{2} D_{\alpha}^{0} \mathcal{N}_{\alpha}(\epsilon_{F})$. 
Here $D_{\alpha}^{0}$ and $\mathcal{N}_{\alpha}(\epsilon_{F})$ denote the bare diffusion coefficient and the density of states at the Fermi level for spin $\alpha$, respectively. 
With the band spin splitting neglected, $\mathcal{N}_{\alpha}(\epsilon_{F}) = \tfrac{1}{2}\mathcal{N}_{\mathrm{e}}(\epsilon_{F})$.
} & $0.32$ $(\mu\Omega\,\text{cm})^{-1}$  \\Bare electron diffusion constant & $D^0_{\mathrm{s}}$ & $ \overline{\tau_\mathrm{e}} v_F^2 / 3$ & $6.67\times10^{-3}\,\text{m}^2/\text{s}$  \\
        Bare electron diffusion length & $\lambda^0_{\mathrm{s}}$ & $\sqrt{v^2_F\overline{\tau_\mathrm{e}} \tau_{\uparrow \downarrow}/6}$ & $57.7$ nm  \\
        Bulk exchange interaction & $J^0_{\mathrm{sd}}$ & & $0.1$ eV \\
        Curie temperature & $T_C$ & -- & $1400$ K  \\
        Local spin per lattice site  & $S$ & -- & $3/2$  \\Lattice constant & $a_{0,F}$ & -- & $4$ \AA  \\
Magnon stiffness  & $A_{\mathrm{ms}}$ & $2J_{\mathrm{dd}}S a_{0,F}^2$~\footnote{Here $J_{dd}$ denotes the exchange interaction between local moments, which, within mean field theory, scales with the Curie temperature of the ferromagnet as $J_{dd}\sim 3k_BT_C/\pi^2S(S+1)$~\cite{kittel91book-QTSolid,Gubanov87JMMM_LSD-Jex-FM,Pajda01PRB_Tc-SW-FM(theor)}. }& $273.5~\text{meV}\cdot\text{\AA}^2$~\cite{Kuzmin05PRL_magnon-stiff}  \\
        Magnon gap & $\Delta^0_\mathrm{g}$ & -- & $1$ meV  \\
        Magnon conserving relaxation time & $\tau^0_{\mathrm{m}}$ & -- & $10$ ps \\
        Magnon non-conserving relaxation time & $\tau^0_{\mathrm{th}}$ & -- & $100$ ps  \\
        Bare magnon diffusion constant & $D^0_{\mathrm{m}}$ & $ \tau_{\mathrm{m}} \overline{v_q^2} / 3$ & $7.4\times10^{-4}\,\text{m}^2/\text{s}$  \\
        Bare magnon diffusion length & $\lambda^0_{\mathrm{m}}$ & $\sqrt{\overline{v^2_q}\tau_{\mathrm{m}} \tau_{\mathrm{th}}/3}$~\cite{ZZ12PRB_spin-convertance} & $0.3$ $\mu$m  \\
    \end{tabular}
    \end{ruledtabular}
\end{table*}

\begin{table*}
    \centering
    \caption{Parameters for electron spin transport in the NM layer}
    \label{table:II-NM}
    \renewcommand{\arraystretch}{1.5}
    \begin{ruledtabular}
    \begin{tabular}{lccc}
        \textbf{Description} & \textbf{Symbol} & \textbf{Expression} & \textbf{Value} \\
        \hline
        Lattice constant & $a_{0,N}$ & -- & $4$ \AA  \\Electron momentum relaxation time & $\tau_\mathrm{e}$ & - & $0.02$ ps \\
        Spin-Hall angle & $\theta_{\mathrm{sh}}$ & -- & $0.1$~\cite{aHoffman13IEEE_SHE,Sinova15RMP_SHE} \\
        Drude conductivity & $\sigma_{0,N}$ & $e^2 D^0_{\mathrm{s}} \mathcal{N}_\mathrm{e}(\epsilon_F)$ & $0.16$ $(\mu\Omega\,\text{cm})^{-1}$  \\
        Electron spin diffusion length & $\lambda_N$ & -- & $5$ nm~\cite{aHoffman13IEEE_SHE,Sinova15RMP_SHE}  \\
    \end{tabular}
    \end{ruledtabular}
\end{table*}

\begin{table*}
    \centering
    \caption{Parameters for coupled electron–magnon transport at the interface and within the FM layer}
    \label{table:III-FM|NM}
    \renewcommand{\arraystretch}{1.5}
    \begin{ruledtabular}
    \begin{tabular}{l c c c}
        \textbf{Description} & \textbf{Symbol} & \textbf{Expression} & \textbf{Value}~\footnote{All values, except for the given parameter $J_{\mathrm{sd}}^0$, are calculated at $T=300$~K and $B=0$~T.} \\
        \hline
        
        Interfacial exchange interaction & $J^0_{\mathrm{sd}}$ & & $0.1$ eV \\
        Spin convertance  & $G_{\mathrm{em}}$ & $(\pi S/\hbar)J^2_{\mathrm{sd}}\mathcal{N}_\mathrm{e}(\epsilon_F)(a_{0,{F}}a_{0,N})^2 \alpha_\mathrm{em}(T)$~\footnote{$\alpha_{\mathrm{em}}$ is a dimensionless and temperature-dependent coefficient given by $\alpha_{\mathrm{em}}=a^3_{0,F}\int_{\omega}\omega\left[-\partial_{\omega} n^0(\omega)\right]$, where we have introduced the shorthand notations $\partial_{\omega}\equiv \partial/\partial\omega$ and  $\int_{\omega}\equiv\int_{\Delta_\mathrm{g}}^{E_{\mathrm{max}}}\mathrm{d}\omega \mathfrak{g}_{m}(\omega)$ with $\mathfrak{g}_{\mathrm{m}}(\omega)=\int \frac{\mathrm{d}^3 q}{(2\pi)^3}\delta(\omega-\omega_q)$ is the magnon density-of-states at energy $\omega$~\cite{Moos68PRL_magnon-DOS-GdCl3}.}~\cite{ZZ12PRB_spin-convertance} & 4620.8 m/s\\
        Spin convertance & $G_{\mathrm{me}}$ & $(\pi S/\hbar)J^2_{\mathrm{sd}}\mathcal{N}_\mathrm{e}(\epsilon_F)(a_{0,F}a_{0,N})^2 \alpha_\mathrm{me}(T)$~\footnote{$\alpha_{\mathrm{me}}$ is a dimensionless and temperature-dependent coefficient given by $\alpha_{\mathrm{me}}=\mathcal{N}_\mathrm{e}(\epsilon_F)a_{0,F}^3\bar{E}_\mathrm{m}(T)/4$, where $\bar{E}_\mathrm{m}(T)\left[=\int_{\omega}\omega n^0(\omega)/\int_{\omega}n^0(\omega)\right]$ may be regarded as the averaged energy of equilibrium magnons. }~\cite{Takahashi10JPCS_Gem,Bender12PRL_Gem,ZZ12PRB_spin-convertance}  & 176.6 m/s~ \\
        Element of the relaxation-rate matrix  $\boldsymbol{\tau}^{-1}$  & $\tau_{11}$ & Eq.~(\ref{eq:tau_ij}a) & 0.12 ps \\
        Element of the relaxation-rate matrix  $\boldsymbol{\tau}^{-1}$  & $\tau_{12}$ & Eq.~(\ref{eq:tau_ij}b) & 4.76 ps \\
        Element of the relaxation-rate matrix  $\boldsymbol{\tau}^{-1}$  & $\tau_{21}$ & Eq.~(\ref{eq:tau_ij}c) & 0.32 ps \\
        Element of the relaxation-rate matrix  $\boldsymbol{\tau}^{-1}$  & $\tau_{22}$ & Eq.~(\ref{eq:tau_ij}d) & 8.70 ps \\
        Element of the diffusion-coefficient matrix  $\mathbf{D}$ & $D_{\mathrm{s}}$ & Eq.~(\ref{eq:diffusion-matrix-elements}a) & $5.8\times10^{-3} \, \text{m}^2/\text{s}$ \\
         Element of the diffusion-coefficient matrix  $\mathbf{D}$ & $D_{\mathrm{sm}}$ & Eq.~(\ref{eq:diffusion-matrix-elements}b) & $8.8\times10^{-7} \, \text{m}^2/\text{s}$   \\
         Element of the diffusion-coefficient matrix  $\mathbf{D}$ & $D_{\mathrm{ms}}$ & Eq.~(\ref{eq:diffusion-matrix-elements}c) & $8.3\times10^{-6} \, \text{m}^2/\text{s}$ \\
        Element of the diffusion-coefficient matrix  $\mathbf{D}$ & $D_{\mathrm{m}}$ & Eq.~(\ref{eq:diffusion-matrix-elements}d) & $2.9\times10^{-4} \, \text{m}^2/\text{s}$ \\
        Effective electron diffusion length & $\lambda_{\mathrm{s}}$ & Eq.~(\ref{eq:lambda_entries}a) & 26.4 nm \\
        Cross-diffusion length & $\lambda_{\mathrm{sm}}$ & Eq.~(\ref{eq:lambda_entries}b) & 166.1 nm \\
        Cross-diffusion length & $\lambda_{\mathrm{ms}}$ & Eq.~(\ref{eq:lambda_entries}c) & 9.6 nm  \\
        Effective magnon diffusion length & $\lambda_{\mathrm{m}}$ & Eq.~(\ref{eq:lambda_entries}d) & 50.2 nm \\  
    \end{tabular}
    \end{ruledtabular}
\end{table*}

\section{Expressions for the diffusion-coefficient and relaxation-rate matrices}
\label{appendix:D&tau matrices}

The derivation of Eqs.~(\ref{eq:current-continuity and drift-diffusion}) and (\ref{eq:uc-continuity-eqs}) was originally presented in Ref.~\cite{yhCheng17PRB_magnon-el}. Here we rederive these equations, outlining only the key steps for brevity, and then provide generalized integral expressions for the entries of the diffusion and scattering-rate matrices (i.e.,~\(\mathbf{D}\) and \(\boldsymbol{\tau}^{-1}\)),
 which enable exploration of a broader parameter space beyond the approximations used in Ref.~\cite{yhCheng17PRB_magnon-el}.
 
We starts with a linearization of the coupled kinetic equations. The electron and magnon distribution functions are written as an equilibrium piece plus a small nonequilibrium deviation \begin{subequations}
\begin{align}
\label{eq:electron_ansatz}
    f_{\mathbf{k}\sigma}(\mathbf{r}) &= f^0(\epsilon_{k}) - \frac{\partial f^0(\epsilon_{k})}{\partial \epsilon_k} 
\left[ \delta \mu_\sigma(\mathbf{r}) + g_{\mathbf{k}\sigma}(\mathbf{r}) \right],
\\
\label{eq:magnon_ansatz}
n_\mathbf{q}(\mathbf{r}) &= n^0(\omega_q) - \frac{\partial n^0(\omega_q)}{\partial \omega_q} 
\left[ \delta \mu_{\mathrm{m}}(\mathbf{r}) + g_{\mathbf{q}\mathrm{m}}(\mathbf{r}) \right],
\end{align}
\end{subequations}
where $f^0(\epsilon_{k})=\{\exp\left[(\epsilon_k-\epsilon_F)/k_BT\right]+1\}^{-1}$ and $n^0(\omega_q)=\left[\exp(\omega_q/k_BT)-1\right]^{-1}$ are the equilibrium Fermi–Dirac and Bose–Einstein distribution functions of electrons and magnons, respectively. The deviations from equilibrium are decomposed into isotropic parts $\delta\mu_{\alpha}$ and anisotropic parts $g_{\mathbf{p}\alpha}(z)$, where $\mathbf p=\mathbf k$ for electrons ($\alpha=\sigma$) and $\mathbf p=\mathbf q$ for magnons ($\alpha=m$). The anisotropic part satisfies $\frac{1}{4\pi}\int d\Omega_p\, g_{\mathbf{p}\alpha}=0$, where $\Omega_p$ denotes the solid-angle measure over directions in momentum space. For simplicity, we assume the NM and FM layers share the same Fermi energy $\epsilon_F$ and a single parabolic conduction band with identical effective mass $m$. However, the analysis can be readily extended to cases where the conduction bands of the FM are spin split, and where $m$ and $\epsilon_F$ differ between the NM and FM, while the essential physics remains intact.

Note that although magnon number is not strictly conserved, a local magnon chemical potential \(\mu_\mathrm{m}\) can still be defined when
number–conserving equilibration (associated with relaxation time \(\tau_\mathrm{m}\)) is much faster than
number–nonconserving relaxation (with \(\tau_{\mathrm{th}}\)), i.e.
$
\tau_\mathrm{m} \ll \tau_{\mathrm{th}}$ 
so that magnon number is approximately conserved on the relevant time scales.
This is analogous to the case of electron spin transport, where a spin chemical potential \(\mu_{\sigma}\) is meaningful if the momentum–relaxation time of electrons with spin $\sigma$ is much shorter than the spin–flip time, i.e., 
$\tau_{\mathrm{\sigma}} \ll \tau_{\mathrm{\uparrow\downarrow}}$~\cite{Wyder87PRL_rb-NM-FM,Fert&Valet93PRB_cpp-GMR}.
In addition, a finite magnon gap  may ensure 
so the expansion
is controlled for small \(|\mu_\mathrm{m}|\).

The expansion of the distribution functions and the resulting linearization of the collision terms in the kinetic equations are valid provided that the spin-dependent electron and magnon chemical potentials satisfy
\[
\delta\mu_{\sigma},\, \delta\mu_{\mathrm m} \ll k_B T \ll \epsilon_F,\, E_{\max},
\]
where \(\delta\mu_{\sigma}\) denotes the spin-dependent electron chemical potential, \(\delta\mu_{\mathrm m}\) the magnon chemical potential, \(\epsilon_F\) the electronic Fermi energy, and \(E_{\max}\) the maximum magnon energy, which is associated with the Curie temperature at the mean-field level (i.e., \(E_{\max}\sim k_B T_C\)). This condition ensures the validity of the linearization of the Fermi--Dirac and Bose--Einstein distributions around locally equilibrated distributions characterized by position-dependent chemical potentials, as well as the existence of well-defined quasiparticles. The anisotropic corrections \(g_{\mathbf{k}\sigma}\) and \(g_{\mathbf{q}\mathrm m}\), which encode the transport-induced distortions of the distribution functions, are likewise treated perturbatively and retained only to leading order, consistent with weak driving and near-equilibrium conditions \cite{sZhang12PRL,Goodings63PR_resistivity-el-magnon,yhCheng17PRB_magnon-el,Fert69JPhysC_two-current-magnon,ZZ12PRB_spin-convertance}.

The linearization of Eqs.~(\ref{eq:KE-electron}) and (\ref{eq:KE-magnon}) is achieved by substituting the ansatz, (\ref{eq:electron_ansatz}) and (\ref{eq:magnon_ansatz}), and retaining only terms first order in $\mu_\alpha$ and $g_\alpha$ ($\alpha=\sigma~\text{or}~m$). The macroscopic transport equations, (\ref{eq:current-continuity}) and (\ref{eq:drift-diffusion}), then follow from the zeroth and first velocity moments of the linearized kinetic equations: Multiplying the electron (magnon) equation by $\mathbf{v}_{\mathbf{k}\sigma}^n$ ($\mathbf{v}_{\mathbf{q}\mathrm{m}}^n$) with the moment index $n=0,1$, perform the integration over the corresponding momentum space. The $n=0$ moment yield the generalized continuity equations~(\ref{eq:current-continuity}), while $n=1$ moment gives the constitutive drift-diffusion (Ohim-Fick) relations~(\ref{eq:drift-diffusion}), linking the currents to external electric field and gradients of carrier densities. 

The elements of the relaxation-rate matrix, i.e., $\boldsymbol{\tau}^{-1}$ in Eq.~(\ref{eq:current-continuity}), can be expressed as
\begin{subequations}\label{eq:tau_ij}
\begin{align}
\tau^{-1}_{11} &= 2\,\left(\tau_{\uparrow\downarrow}^{-1}
 + \tau_T\,\tau_{J}^{-2}\right),\\
\tau^{-1}_{12} &= \tau_T^{\prime}\,\tau_{J}^{-2},\\
\tau^{-1}_{21} &= \tau_T\,\tau_{J}^{-2},\\
\tau^{-1}_{22} &= \tau_{\mathrm{th}}^{-1}
 + \frac{1}{2}\tau_T^{\prime}\,\tau_{J}^{-2}.
\end{align}
\end{subequations}
where $\tau_{J}\equiv \hbar/J_\mathrm{sd}$, and the two characteristic time scales, $\tau_T$ and $\tau_T^{\prime}$, have temperature dependences implicit in the equations below:
\begin{subequations}
\begin{align}
\tau_T&=\frac{\hbar\,S(k_Fq_C^2a_0^3)}{ 4 \pi k_BT}\cdot\mathcal{I}_T^{(1,1)}\\
\tau_T^{\prime}&=\tau_T\cdot\frac{\mathcal{N}_\mathrm{e}(\epsilon_F)}{\mathcal{N}_\mathrm{m}(\Delta_\mathrm{g}^{\mathrm{eff}})}
\end{align}
\end{subequations}
wherein $\mathcal{N}_\mathrm{e}(\varepsilon_F)(\approx\frac{mk_F}{\pi^2\hbar^2})$ is the three-dimensional electron density of states (DOS) per unit energy per unit volume at the Fermi level, and
\begin{equation}
\mathcal{N}_\mathrm{m}(\Delta^{\text{eff}}_\mathrm{g})=\int \frac{d^3q}{(2\pi)^3} \left(-\frac{\partial n^0(\omega_ q)}{\partial \omega_q}\right)\,,
\end{equation}
which serves as a thermally weighted magnon spectral factor—distinct from the magnon DOS—and quantifies the density of thermally active magnons within an energy window of order 
$k_B T$ above the effective gap.   $\Delta^{\mathrm{eff}}_\mathrm{g}\,(\equiv \Delta_\mathrm{g}+g\mu_B \mathbf{m}\cdot\mathbf{B})$, i.e., the minimum energy required to excite a magnon.  $\mathcal{I}^{(1,1)}_T$ is a dimensionless function of temperature; it's explicit---and generalized---form, together with three related integral functions, will be presented collectively in Eq.~(\ref{eq:tInts}) for ease of comparison.

From the expressions of the relaxation-rate matrix given by Eq.~(\ref{eq:tau_ij}), it is clear that in the absence of electron-magnon scattering (i.e., $\tau_{J}\to \infty$), the off-diagonal elements of the  $\boldsymbol{\tau}^{-1}$ matrix vanish, and the continuity equations for electrons and magnons reduce to their uncoupled forms [cf.~Eqs~(\ref{eq:uc-continuity-eqs}a) and (\ref{eq:uc-continuity-eqs}b)].

For the diffusion-coefficient matrix in Eq.~(\ref{eq:drift-diffusion}), it is convenient (notation-wise) to parametrize it as $\mathbf{D}=\boldsymbol{\eta}^{-1}$, and relate the elements of two matrices via
\begin{subequations}
\label{eq:diffusion-matrix-elements}
\begin{align}
D_\mathrm{s} &= \frac{\eta_{22}}{\det(\boldsymbol{\eta})}\,,\\
D_\mathrm{sm} &= \frac{\eta_{12}}{\det(\boldsymbol{\eta})}\,, \\
D_\mathrm{ms} &= \frac{\eta_{21}}{\det(\boldsymbol{\eta})}\,, \\
D_\mathrm{m} &= \frac{\eta_{11}}{\det(\boldsymbol{\eta})}
\end{align}
\end{subequations}
with the elements of the $\boldsymbol{\eta}$ matrix given by 
\begin{subequations}
\begin{align}
    \eta_{11}&=\frac{3}{v^2_F} \left( \bar{\tau}^{-1}_\mathrm{e}+\tau^{-1}_{\uparrow\downarrow}+\tau^{-2}_J \,\tau_{T,1} \right) 
    \\
    \eta_{12}&=\frac{3}{v^2_F}\left( \tau^{-2}_J \, \tau_{T,2}\right)
    \\ 
    \eta_{21}&=\frac{3}{\overline{v_q^2}}\left(\tau^{-2}_J \, \tau_{T,3}\right)
    \\
    \eta_{22}&=\frac{3}{\overline{v_q^2}}\left(\tau^{-1}_\mathrm{m}+\tau^{-1}_{\mathrm{th}}+\tau^{-2}_J\,\tau_{T,4}\right)
\end{align}
\end{subequations}
Here $\bar{\tau}_\mathrm{e}[=2\tau_{\uparrow}\tau_{\downarrow}/(\tau_{\uparrow}+\tau_{\downarrow})]$ is the harmonic mean of the electron momentum relaxation time. $\tau_{T,i}$ ($i=1,2,3,4$) are four temperature dependent time scales, with explicit expressions given by 
\begin{subequations}
\begin{align}
\tau_{T,1}
&= \frac{S\hbar\, (k_F q_C^2 a_0^3)}{4\pi k_B T}
\left\{\mathcal{I}_T^{(2,1)} + \mathcal{J}_T^{(2,1)}+\left(\frac{q_C}{2k_F}\right)^2 \right. \nonumber\\
 &\quad\times \left. \left[\mathcal{K}_T^{(0,3)}-\mathcal{I}_T^{(0,3)}+3\left(\mathcal{L}_T^{(0,3)} + \mathcal{J}_T^{(0,3)}\right)\right]\right\}\\
\tau_{T,2}&=\frac{S\hbar\, (q_C a_0)^3}{8\pi k_B T}\cdot\frac{\mathcal{N}_e(\epsilon_F)v_F }{\mathcal{N}_m(\Delta_g^{\mathrm{eff}})\overline{v_q}}\cdot\mathcal{K}_T^{(1,2)}\\
\tau_{T,3}&=\frac{S\hbar (q_Ca_0)^3}{16\pi k_BT}\left(\frac{v_C}{v_F}\right)\nonumber\nonumber\\
&\quad\times\left(\mathcal{K}_T^{(0,3)}-\mathcal{I}_T^{(0,3)}+\mathcal{L}_T^{(0,3)}+\mathcal{J}_T^{(0,3)}\right)\\
\tau_{T,4}&=\frac{S\hbar (k_Fq^2_Ca^3_0)}{8\pi k_BT}\cdot\frac{\mathcal{N}_e(\epsilon_F)v_C}{\mathcal{N}_m(\Delta_g^{\mathrm{eff}})\overline{v_q}}\cdot \mathcal{I}_T^{(1,2)}
\end{align}
\end{subequations}
where $v_F$ is the Fermi velocity. $v_C(\equiv 2A_\mathrm{ms}q_C$) can be regarded as the maximum magnon group velocity, where  $q_C\equiv\sqrt{(E_{\max}-\Delta_\mathrm{g})/A_{\mathrm{ms}}}$ with the maximum magnon energy set by the Curie temperature of the FM via $E_{\max}\approx 3k_B T_C/(S+1)$ in the mean field approximation~\cite{Gubanov87JMMM_LSD-Jex-FM,Pajda01PRB_Tc-SW-FM(theor),sZhang97PRL-magnon-hot-el-MTJ}. We have also defined the mean magnon speed and the mean-squared magnon speed as
\begin{subequations}
\begin{align}
\overline{v_q}&=\frac{\int \frac{d^3q}{(2\pi)^3} \, \left[-\frac{\partial n^0(\omega_q)}{\partial \omega_q}\right]v_q}{\int \frac{d^3q}{(2\pi)^3} \left[-\frac{\partial n^0(\omega_ q)}{\partial \omega_q}\right]}\\
\overline{v_q^2}&=\frac{\int \frac{d^3q}{(2\pi)^3} \left[ - \frac{\partial n^0(\omega_{q})}{\partial \omega_q} \right]v^2_{q} }{\int \frac{d^3q}{(2\pi)^3} \left[-\frac{\partial n^0(\omega_ q)}{\partial \omega_q}\right]}
\end{align}
\end{subequations}
Note that the temperature dependencies of $\tau_{T,i}$ ($i=1,2,3,4$) are governed by specific electron-magnon scattering processes and are embodied in the following dimensionless integral functions:
\begin{subequations}\label{eq:tInts}
\begin{align}
\mathcal{I}_T^{(s,t)}&=\iint \mathrm{d}\tilde{k}\,\mathrm{d}\tilde{q}\,\tilde{k}^s\,\tilde{q}^t\,\nonumber\\&\quad\quad\times n^0(\omega_q)\,f^0(\epsilon_k)\,\left[1-f^0(\epsilon_k+\omega_q)\right]\,\\
\mathcal{J}_T^{(s,t)}&=\iint \mathrm{d}\tilde{k}\,\mathrm{d}\tilde{q}\,\tilde{k}^s\,\tilde{q}^t\,\nonumber\\&\quad\quad\times n^0(\omega_q)\,f^0(\epsilon_k-\omega_q)\,\left[1-f^0(\epsilon_k)\right]\,\\
\mathcal{K}_T^{(s,t)}&=\iint \mathrm{d}\tilde{k}\,\mathrm{d}\tilde{q}\,\tilde{k}^s\,\tilde{q}^t\,\left(\frac{\omega_q}{\epsilon_q}\right)\nonumber\\&\quad\quad\times n^0(\omega_q)\,f^0(\epsilon_k)\,\left[1-f^0(\epsilon_k+\omega_q)\right]\,\\
\mathcal{L}_T^{(s,t)}&=\iint \mathrm{d}\tilde{k}\,\mathrm{d}\tilde{q}\,\tilde{k}^s\,\tilde{q}^t\,\left(\frac{\omega_q}{\epsilon_q}\right)\nonumber\\&\quad\quad\times n^0(\omega_q)\,f^0(\epsilon_k-\omega_q)\,\left[1-f^0(\epsilon_k)\right]\,
\end{align}
\end{subequations}
where $\tilde{k}=k/k_F$, $\tilde{q}= q/q_C$, and $s,t$ are integer power indices. When evaluating the \(q\)-integrals, we impose an ultraviolet cutoff \(q_{\max}=q_C\).

\section{Dominance of Short-Wavelength (High-\(q\)) Magnons in UMR}

\label{appendix:Dipolar-interaction}

\begin{figure}[htbp]  \includegraphics[width=0.45\textwidth]{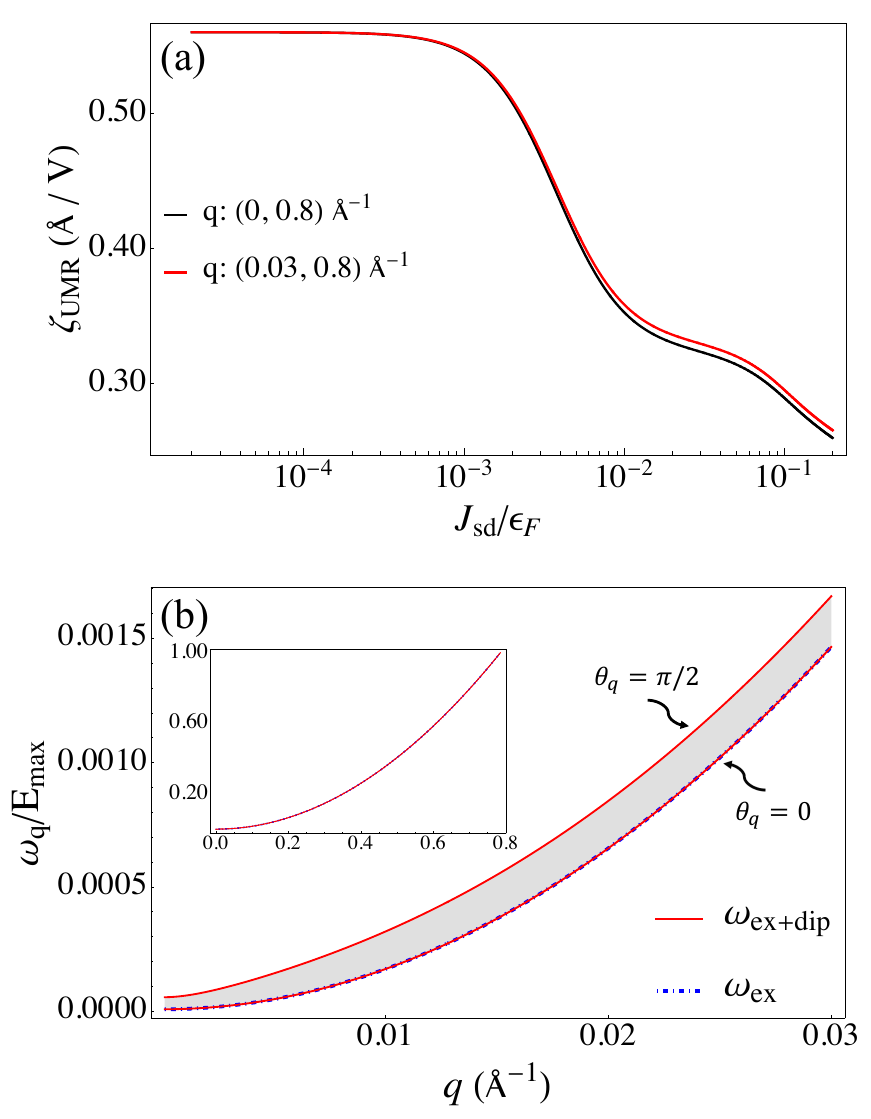}
  \caption{(a) UMR coefficient $\zeta_{\mathrm{UMR}}$ as a function of the
exchange coupling $J_{\mathrm{sd}}/\epsilon_F$, computed using different wave-vector integration windows. The black curve includes contributions from the full magnon spectrum $(q = 0\text{--}0.8~\mathrm{\AA}^{-1})$, while the red curve excludes long-wavelength
magnons $(q < 0.03~\mathrm{\AA}^{-1})$, where dipolar interactions are relevant. (b) Magnon dispersion illustrating the relative roles of exchange and dipolar interactions.
The magnon energy $\omega_q$, normalized by the maximum magnon energy $E_{\max}$, is plotted as a function of wave vector $q$ in the long-wavelength regime. Solid red curves show the full exchange--dipolar dispersion $\omega_{\mathrm{ex+dip}}$, while the blue dashed curve denotes
the pure exchange contribution $\omega_{\mathrm{ex}}$. Results are shown for two propagation angles $\theta_q$ relative to the magnetization direction. For $\theta_q = 0$, the full dispersion coincides with the exchange-only result because no magnetic charge is generated, despite the presence of dipolar interactions; in contrast, for $\theta_q = \pi/2$, dipolar interactions significantly enhance the magnon energy at small $q$. The inset displays the dispersion over the full wave-vector range, where all curves merge, demonstrating that exchange interactions dominate at large $q$ and that dipolar effects are confined to the long-wavelength limit. }
\label{fig:dipolar-dispersion}
\end{figure}

In our calculations of magnon-mediated transport quantities, we integrate over the full range
of magnon wave vectors \(q \in [0, q_{\max}]\). Physically, however, the contribution to the UMR
is dominated by short-wavelength (high-\(q\)) magnons. As illustrated in Fig.~\ref{fig:dipolar-dispersion}a, long-wavelength
magnons predominantly induce forward scattering and therefore contribute much less to momentum
relaxation than short-wavelength magnons, which are most effective in relaxing the electron
momentum and thus controlling the UMR.

Correspondingly, the magnons that dominate the UMR lie in the high-\(q\) regime, where the magnon
dispersion is governed primarily by exchange interactions. While long-range magnetostatic
(dipolar) interactions renormalize the magnon spectrum at small wave vectors and introduce an
angular dependence with respect to the magnetization direction~\cite{Holstein40PR_HP-transformations},
their effect becomes negligible at larger wave vectors, where the quadratic exchange term
proportional to \(q^{2}\) prevails. This is demonstrated explicitly in Fig.~\ref{fig:dipolar-dispersion}b, which compares the
exchange-only magnon dispersion with the full dispersion including both exchange and dipolar
interactions: over the wave-vector range relevant for transport, the two dispersions essentially
overlap. This observation justifies the use of the exchange-only magnon dispersion [Eq.~(\ref{Eq:magnon-disp})] in
our calculations and confirms that neglecting dipolar interactions is a controlled and
well-justified approximation for the magnon-mediated UMR considered in this work.

\bibliography{Bilayer-UMR-v2}

\end{document}